
  \input amsppt.sty
  \nologo \NoBlackBoxes
  \magnification=\magstep1


  \topmatter
  \title Asymptotic Solutions to the
 Knizhnik-Zamolodchikov Equation and Crystal
 Base\endtitle
  \author A. N. Varchenko$^\ast$ \endauthor
\dedicatory Dedicated to the memory of Ansgar Schnizer \enddedicatory
\date March, 1994 \enddate
 \thanks \flushpar The author was supported by NSF Grant
 DMS-9203929 \newline
 $^\ast$Department of Mathematics, University of North
 Carolina, Chapel Hill, NC  27599, USA \newline
 email address: {\it varchenko\@math.unc.edu}\endthanks
\abstract
The Knizhnik-Zamolodchikov equation associated with
$s\ell_2$ is considered. The transition functions between
 asymptotic solutions to the Knizhnik-Zamolodchikov
 equation are described. A connection between asymptotic
 solutions and the crystal base in the tensor product of
 modules over the quantum group
$U_qs\ell_2$ is established, in particular, a correspondence between
the Bethe vectors of the Gaudin model of an inhomogenious magnetic chain
and the $\Bbb Q-$basis of the crystal base.
\endabstract
  \leftheadtext{A. Varchenko}
  \rightheadtext{Asymptotic Solutions to the K-Z Equation and
 Crystal Base}
  \endtopmatter
 \document

\head {Introduction} \endhead

 In this work we describe transition functions between
 asymptotic solutions to the Knizhnik-Zamolodchikov (KZ)
 equation and establish a connection between asymptotic
 solutions and the crystal base in the tensor product of
 modules over a quantum group.

 We consider the KZ equation associated with $s\ell_2$
 and the quantum group \newline
$U_qs\ell_2$ , general case can be
considered similarly.

 For a positive integer $m$, denote by $L(m)$ the
 $s\ell_2$ irreducible module with highest weight $m$.
 For positive integers $m_1,\dots,m_n $, set
 $L=L(m_1)\otimes\dots\otimes L(m_n)$.

 Let $\Omega = \frac12 h\otimes h+e\otimes f+f\otimes
 e\in s\ell_2^{\otimes 2}$ be the Casimir operator.  For
 $i\neq j$ denote by $\Omega_{ij}$ the linear operator on
 $L$ which acts as $\Omega$ on the $i$-th and $j$-th
 factors and as the identity on the other factors.  The KZ
 equation on an $L$-valued function $\psi(z_1,\dots
 ,z_n)$ is the system of equations
 $$
 \frac{\partial\psi}{\partial z_j} =
  \frac{1}{\kappa} \sum_{\ell \neq j}
 \frac{\Omega_{j\ell}}{z_j-z_\ell} \psi ,
\qquad j = 1, \dots , n ,
 $$
 where $\kappa$ is a complex parameter.  In this
 paper we assume that $\kappa$  is not a rational
 number.  The KZ equation is defined over $\Cal U_n =\{
 z\in \Bbb C^n\,\vert\, z_i\neq z_j\text{ for }i\neq
 j\}$.

 For any permutation $\sigma\in S_n$, define an open
 convex unbounded polytope
$$
D _{\sigma} = \{ z\in
 \Bbb R^n\,\vert\, z_{\sigma(1)}<\dots <
 z_{\sigma(n)}\}.
$$

 One can easily resolve singularities of $D
 _{\sigma}$ and cover $D _{\sigma}$ by local
 charts with local coordinates $u_1,\dots,u_n$ such that:
 \roster
 \item "(0.1)" $u_n= z_1+\dots +z_n$ , \endroster
 \vskip.5ex
 \roster
 \item "(0.2)" $D _{\sigma}$ is defined in this
 chart by inequalities $u_1>0,\dots, u_{n-1}>0$ ,
 \endroster
 \vskip.5ex
 \roster
 \item "(0.3)" The KZ equation in this chart has the
 form
 $$
 \frac{\partial\psi}{\partial u_n} = 0,\
 \frac{\partial\psi}{\partial u_j} = \frac{1}{\kappa}
 H_j,
 \qquad j=1,\dots,1-n,
 $$
 where $H_j= \Omega_j/u_j+\text{Reg}_j,\ \Omega_j$ is a
 constant operator on $L$, and Reg$_j$ is an operator
 regular at $u=0$.
\endroster

 We call such charts {\it asymptotic zones,} they are numerated
 by suitable trees $T$.

 The operators $\Omega_1,\dots,\Omega_{n-1}$
 commute and have a common eigenbasis.  We distinguish an
 eigenbasis $\Cal B= (v_1(\kappa),\dots, v_N(\kappa))$ for
 nonrational $\kappa$.  For every $j$ the vector $v_j(\kappa)$ is
 proportional to a fixed common eigenvector of
 $\Omega_1,\dots,\Omega_{n-1}$, and the coefficient of
 the proportionality appropriately depends on $\kappa$, see (2.3.1).

 For each $j=1,\dots, N$ and a nonrational  $\kappa$, there exists a unique
 solution $\psi_j$ to the KZ equation with parameter $\kappa$
 defined in $D_{\sigma}$ and
 such that
 $$
 \psi_j = (v_j(\kappa) + \Cal O(u, \kappa)) \prod^{n-1}_{\ell =1}
 u_\ell ^{\mu_{\ell }/\kappa}
 \tag0.4
 $$
 where $\mu_\ell $ is the eigenvalue of $\Omega_\ell $
 at $v_j$, $\Cal O(u, \kappa)$ is a regular function of $u$ at $u=0$ , and
$\Cal O(0, \kappa)=0$. ( Notice that all numbers $\mu_l$ are real. )

 Thus, for each asymptotic zone $( T,\sigma )$ we
 construct a fundamental system of solutions
 $\psi_{T,\sigma}= \{\psi_j\}$.
\bigskip

 Our first main result gives explicit transition
 functions between these fundamental solutions
$\psi_{T,\sigma}$ and $\psi_{T',\sigma '}$ in terms of the
 $q$-$6j$-symbols where $q=q(\kappa)$ and $q(\kappa)= \exp
 (2\pi i/\kappa)$, see (2.4).
\bigskip

 In particular,  if $\kappa=is,\ s\in \Bbb R,\ s\rightarrow
 +0$, then $q(\kappa)\to \infty$, and our formulae show that
 for every $\sigma,\ T,\ T'$ the transition function
 between the fundamental systems of solutions
 $\psi_{T,\sigma}$ and $\psi_{T',\sigma}$ has the form
 $\boldkey 1+\Cal O(q(\kappa)^{-\frac{1}{4}})$, where
 $\boldkey 1$ is the unit matrix and the solutions
 composing $\psi_{T,\sigma}$ and $\psi_{T',\sigma}$  must
 be suitably enumerated, see (2.4) and (1.3.12).

 Assume that $\kappa=is,\ s\in \Bbb R$ and $s\to +0$.  For any
 asymptotic zone consider the fundamental system of solutions
 defined by (0.4).  Our second main result, see (2.5),
 states that for every $j$ the solution $\psi_j$ has an
 asymptotic expansion of the form
 $$
 \psi_j =
 \prod^{n-1}_{\ell =1}
 u_\ell ^{-i\mu_\ell/s}
 \exp(-iS/s)
 \sum^{\infty}_{\ell =0} f_\ell s^\ell
 \tag0.5
 $$
 where $S$, $f_\ell $ are real-analytic functions of $u$ in our
 chart regular at $u=0$, the function $S$ is real
 valued, the functions $f_\ell $ are $L$-valued, and, in
 particular,
$$
\big\vert \prod_{\ell = 1} ^{n-1} u_\ell^{-i\mu_\ell /s}
 \exp(-iS/s) \big\vert =1 .
$$
  The first term $f_0$ is a
 common eigenvector of the operators $H_1,\dots
 ,H_{n-1}$.  The first terms of asymptotics of the
 solutions {$\psi_j$} form an eigenbasis of the operators
 $H_1,\dots,H_{n-1}$.  We describe its limit at $u\to
 0$.

 If the KZ equation has an asymptotic solution of the
 form
 $$
 \psi(u) =\exp (P(u)/\kappa)\sum_{\ell = 0} ^{\infty}
 g_\ell (u) \cdot  \kappa ^\ell ,
 $$
 one can show that $g_0(u)$ must be a common eigenvector
 of $H_1,\dots,H_{n-1}$ and the function $g_0(u)$ is
 determined up to multiplication by a constant \cite{RV}.
 One can deduce from this remark that a solution having
 an asymptotic expansion of the form described in (0.5)
 is unique up to multiplication by a
 function of $s$ which has the form $1+\Cal O(s)$ where
$\Cal O(0) = 0 $.

 Our two main results have the following quantum group
 interpretation.
 \vskip1ex

 Let $U_q$ be the quantum group corresponding to the
 $s\ell_2$ and $U_{q=q(\kappa)}$ its specialization at $q=q(\kappa)$.
 For a positive integer $m$ denote by $L(m,q)$ (resp. \newline
 $L(m,\ q=q(\kappa))$) the irreducible $U_q$ (resp.
 $U_{q=q(\kappa)}$) module with highest weight $m$.  For
 positive integers $m_1,\dots,m_n $ and a permutation
 $\sigma$ denote by $L^{\sigma}(q)$ the product
 $L(m_{\sigma(1)},q) \otimes\dots\otimes
 L(m_{\sigma(n)},q)$.  Denote by $L^{\sigma}(q=q(\kappa))$ the
 corresponding product for $U_{q=q(\kappa)}$.  Denote by
 Sol$\,(\kappa)_{\sigma}$ the space of solutions over $D
 _\sigma$ to the KZ equation with parameter $\kappa$.

 Our formulae for transition functions between the
 constructed fundamental system of solutions allows us to
 construct an isomorphism
 $$
\pi_\sigma (\kappa) : L^\sigma(q=q(\kappa)) \cong
 \text{Sol}\,(\kappa)_\sigma
 $$
 and show, once again, that the monodromy
 representation of the KZ equation is isomorphic to the
 $R$-matrix representation of the braid group, where $R$
 is the universal $R$-matrix of $U_{q=q(\kappa)}$ ,
 see [K, Dr, V1, FW].  Under the isomorphism
 $\pi_{\sigma}(\kappa)$ each of the constructed fundamental
 systems of solutions $\psi_{T,\sigma}$ gives a basis
 $\Cal B_{T,\sigma}(\kappa)$ in $L^{\sigma}(q=q(\kappa))$, see
 (2.4).

In \cite{Ka} the notion of crystal base is
 defined.  The module $L(m,q)$ has a standard crystal
 base when $q\to \infty$.  The tensor product of the
 standard crystal bases gives a {\it distinguished crystal
 base} in $L^{\sigma}(q)$.  Our isomorphism
 $\pi_{\sigma}(\kappa)$ has the following property:
\roster
\item"{}" {\it For every $T$ ,
 the fundamental system of solutions $\psi_{T,\sigma}$
 lifts to the same distinguished crystal base of
 $L^{\sigma}(q)$, see (2.6).}
\endroster
  Therefore, we may conclude
 that the distinguished crystal base is
given by (0.4) and normalized by property (0.5). This gives an \lq\lq
 asymptotic" definition of the crystal base.
\bigskip

 The proof of transition formulae between the fundamental
 solutions and the proof of property (0.5) are based on
 integral representations for solutions to the KZ
 equation, see [SV, V1].  Any solution in $
 \text{Sol}\, (\kappa)_\sigma$ can be represented
as a linear combination of solutions of the form
 $$
 \psi(z) = \int_{\gamma (z)}\,\Phi \,M(t,z)
 \tag0.6
 $$
 where $M$ is a rational $L$-valued differential form of $z$ and
 some variable $t=(t_1,\dots,t_k ),\ \gamma (z)$ is a
 family of cycles in $t$-space,
 $$
 \Phi(t,z) =
 \prod_{1\leq \ell <j\leq n}
 (z_j-z_\ell )^{m_\ell m_j/2\kappa}
 \prod_{1\leq \ell <j\leq k}
 (t_\ell -t_j)^{2/\kappa}
 \prod^k_{\ell =1}\ \prod^n_{j=1}
 (z_j-t_\ell )^{-m_j/\kappa}.
 $$
 In this construction cycles $\gamma (z)$ numerate solutions to the KZ
 equation.  In \cite{V1} we identified the homology group
 of such cycles for a fixed $k$
with the subspace of singular vectors in
 $L^\sigma(q=q(\kappa))$ of a suitable weight, see precise statements
in \newline [V1] .  This construction after suitable
 renormalization gives us the isomorphism $\pi_{\sigma}$,
 see (4.4).

 If $\kappa =is,\ s\to +0$, then integral (0.6) is localized at
critical points of the function $\Phi$ with respect to the variable
$t$ , and
one can compute asymptotics of
 the solution applying the method of
 steepest descent, see [RV,V2].  One can push the cycle $\gamma (z)$
 onto $t-$critical points of $\Phi$ .
 If $\gamma (z)$ \lq\lq sits" on some of the
 $t$-critical points then each of the points gives its
 input into the asymptotic expansion of the solution.

 The symmetric group $S_k$ of permutations of coordinates
  $t_1,\dots,t_k$ acts on the set of critical points of
 $\Phi$.

 In each asymptotic zone $(T,\sigma)$, one can describe
 asymptotics of $t$-critical points of $\Phi$.  It turns
 out that the $S_k$ orbits of $t$-critical points of
 $\Phi$ are in correspondence with solutions
 $\psi_j$ composing the fundamental system
 $\psi_{T,\sigma}$.  For any $j$ we distinguish a cycle
 $\gamma _j(z)$ such that
\roster
\item $\psi_j= \int_{\gamma
 _j(z)}\Phi M ,$
\item  $\gamma _j(z)$ sits exactly
 on one $S_k$ orbit of $t$-critical points,
\item the
 asymptotic expansion of the integral at this orbit has
 property (0.5).
\endroster

 This gives a \lq\lq topological" definition of the
 crystal base as the base in the homology group of cycles
corresponding to a fixed $z$. This base is
formed by
 the classes numerated by the orbits of $t$-critical points and
each such a  class has a representing chain sitting on one orbit.
In this construction $z$ must be in an asymptotic zone.
\bigskip

The value of the function $M$ at a $t-$critical point $(t(z),z)$
of the function $\Phi$ gives the first term of asymptotics as
$k=is$ and $s \to +0$. This value $M(t(z),z)$ is an eigenvector of
the commuting operators $H_1(z), ... , H_{n-1}(z)$ .

The algebraic Bethe Ansatz is a construction of eigenvectors for a
system of commuting operators.  One considers a
vector valued function of a special form and determines its arguments in
such a way that the value of this function
is an eigenvector. The equations which determine these special values
of arguments are called the Bethe equations. The eigenvector is called
a Bethe vector, for more details see [FT].

One of the systems of commuting operators which can be diagonalized
by the ABA is the Gaudin model of an inhomogeneous
magnetic chain [G].

It turns out that the function $M(t,z)$ is exactly
that special function which appear in the ABA for the Gaudin model and
the Bethe equations for the Gaudin model coincide with the equations
on $t-$critical points of the function $\Phi$, see [B,BF,RV].

This relation to the ABA and the above formulated results give us a
correspondence between the $\Bbb Q-$basis of the distiguished crystal base
and the Bethe vectors of the Gaudin model, see (2.6.5), or , more generally,
between the crystalization phenomenon in the theory of quantum groups
and the Bethe ansatz construction in statistical mechanics. On the Bethe
ansatz and the KZ equation see also [FFR].
\bigskip

 Section 1 contains preliminary information on bases in
 $L$ and $L^\sigma(q)$.  In section 2 we state the
 formulae for transition functions and the theorem about
 asymptotic expansion (0.5).  In section 3 we review
 integral representations for solutions and describe
 asymptotics of critical points and the corresponding
 Bethe vectors.  Section 4 contains
 homological part of the work.
\vskip5ex

\newpage

 \head  1.  Bases in Tensor Products \endhead

 \subhead (1.1)  Quantum group $\bold
 U_{\bold q} \bold{s}\pmb{\ell}
 _{\bold 2}$ \endsubhead

 The $\Bbb Q (q^{1/4})$-algebra $U_q=U_qs\ell_2$ is
 the algebra generated by the symbols $e, \ f,\ h$ with
 the relations
 $$\align
 [h,e]&= 2e, \tag 1.1.1 \\
 [h,f]&= -2f, \\
 [e,f]&= q^{h/2} -q^{-h/2}.
 \endalign
 $$
 Here $q^{1/4}$ is an indeterminate.

  {\bf Remark.} Usually one considers $U_q$ as an $\Bbb Q(q)$ algebra.
Our choice of $q^{1/4}$ instead of $q$ is motivated by topological
considerations in section 4.

  $U_q$ has a
 Hopf algebra structure with the comultiplication $\Delta
 $ defined by
 $$\align
 \Delta (h) &= h\otimes 1 +1 \otimes h, \tag1.1.2 \\
 \Delta (e) &= e\otimes q^{h/2}+ 1\otimes e, \\
 \Delta (f) &= f\otimes 1+q^{-h/2} \otimes f .
 \endalign
 $$
 By $\Delta $, the tensor product of $U_q$-modules has a
 structure of $U_q$-module.

 Let
 $$\align
 &[n]_q = (q^{n/2}-q^{-n/n}) /
 (q^{1/2}-q^{-1/2}),
 \tag  1.1.3\\
 &(n)_q= q^{n/2}-q^{-n/2},\\
 &[n]_q! = \prod^n_{k=1}\,[k]_q,\\
 &e^{(n)}= {e^n \over [n]_q!} ,\quad e^{(0)}=1, \\
 &f^{(n)}= {f^n \over [n]_q!} ,\quad f^{(0)}=1. \\
 \endalign$$

 Let $V_1$ and $V_2$ be $U_q$-modules.  The modules $V_1
 \otimes V_2$ and $V_2\otimes V_1$ are isomorphic.  The
 isomorphism is given by the formula
 $$
 V_1\otimes V_2\overset\text{R}\to{\rightarrow}V_1
  \otimes V_2 \overset\text{P}\to{\rightarrow}V_2
 \otimes V_1
 \tag1.1.4
 $$
 where $P$ is the transposition of the factors and $R\in
 U_q\widehat{\otimes}U_q$ is the universal $R$-matrix .
 Denote the isomorphism $PR$ by $\hat R$.

 Denote by $S_n$ the group of permutations of the set $\{
 1,\dots,n \}$, by $B_n$ the braid group on $n$ strings
 with the standard generators $\sigma_1,\dots,\sigma_n$,
 by $\tau:B_n\to S_n$, the natural epimorphism.

 Let $V_1,\dots,V_n$ be $U_q$-modules.  For $j=1,\dots
 ,n-1$,  define an $U_q$-isomorphism,
 $$
 R_{j,j+1} : V_1 \otimes\dots\otimes V_n \rightarrow
 V_1 \otimes\dots\otimes V_{j+1} \otimes V_j \otimes
\dots\otimes V_n, \tag1.1.5
 $$
 as the map which acts as $\hat R$ on $V_j\otimes V_{j+1}$
 and as the identity on the other factors.

 For any braid $\sigma\in B_n$ these isomorphisms induce
 a well defined isomorphism
 $$
 R_{\sigma} : V_1 \otimes\dots\otimes V_n
 \rightarrow
 V_{\tau\sigma(1)} \otimes\dots\otimes
 V_{\tau\sigma(n)}. \tag1.1.6
 $$
 \vskip5ex

 \subhead (1.2) Bases in Tensor Products \endsubhead

 For $m\in \Bbb Z$ let $V(m)$ be the Verma module  with
 highest weight $m$.  $V(m)$ is generated by its singular
 vector $v_m$ such that $ev_m=0$ and $q^hv_m= q^mv_m$.
 The elements $f^{(j)}v_m,\ j\geq 0$, form a basis of
 $V(m)$.  For a nonnegative $m$ the vector $f^{(m+1)}v_m$
 generates a proper submodule of $V(m)$.  The quotient, $L(m)$, is
 the $(m+1)$-dimensional irreducible $U_q$-module with a
 basis generated by $v_m, \ f^{(1)}v_m,\dots,
 f^{(m)}v_m$.

 The tensor product of irreducible modules is the direct
 sum of irreducible modules:
$$
L(m_1)\otimes L(m_2) =
 L(\vert m_1-m_2 \vert) \oplus L(\vert m_1-m_2 \vert+2)
 \oplus \dots\oplus L(m_1+m_2) ,
$$
and a singular
 vector of $L(m_1+m_2-2k)$ has the form
 $$\multline
 (v_{m_1}\ ,_{k} \ v_{m_2})  =
 \sum^k_{p=0} (-1)^p \
 \frac{(m_2-k+1)_q \dots (m_2-k+p)_q}{(m_1)_q \dots
 (m_1-p+1)_q} \cdot \\
 \vspace{2\jot}
 \cdot q^{-p(m_2-2k+p+1)/2}
 f^{(p)} v_{m_1}\otimes f^{(k-p)} v_{m_2}.
 \endmultline\tag1.2.1
 $$
 This decomposition gives a basis $\{ f^{(\ell)}
 (v_{m_1}\ ,_{k} \ v_{m_2}) \}$ in the tensor product.

 The $U_q$-isomorphism
 $$\gather
 \hat R : L(m_1) \otimes L(m_2) \rightarrow L(m_2) \otimes
 L(m_1)
 \tag1.2.2\\
 \intertext{is given by}
 f^{(\ell)} (v_{m_1}\ ,_{k} \ v_{m_2})
 \mapsto R (m_1,m_2;k)_q f^\ell (v_{m_2}\ ,_{k} \ v_{m_1}),\tag1.2.3\\
  \vspace{2\jot}
 R(m_1,m_2;k)_q = (-1)^k\,
 q^{(m_1m_2-4km_2+2k(k+1))/4}\,
 \frac{(m_2)_q \dots (m_2-k+1)_q }{(m_1)_q \dots
 (m_1-k+1)_q }.
 \endgather
 $$

 There are two ways to decompose $L(m_1) \otimes L(m_2)
 \otimes L(m_3)$ into irreducibles.  These two ways give
 two bases in the triple tensor product: $\{ f^{(\ell)}
 (( v_{m_1},_{k_1} \ v_{m_2}), _{k_2} \ v_{m_3}) \}$ and $\{
 f^{(\ell)} ( v_{m_1},_{k_1} \ (v_{m_2},_{k_2} \ v_{m_3})) \}$.

 The matrix elements of the matrix, connecting these
 bases, are called the $6j$-symbols:
 $$\multline
 f^{(\ell)}
 ((v_{m_1}\ ,_{k_1} \ v_{m_2}),_{k_2} \ v_{m_3})
 = \\ \vspace{3\jot}
\sum_k
 \biggl\{ \matrix m_1,\\k_1,\endmatrix  \matrix m_2,\\k_2,\endmatrix
  \matrix m_3\\k\endmatrix
 \biggr\}_q
 \
 f^{(\ell)}
 (v_{m_1}\ ,_{k} \ (v_{m_2}\ ,_{k_1+k_2-k} \ v_{m_3})) .
 \endmultline\tag1.2.4
 $$

 There are numerous formulae for $6j$-symbols, see,
 for example, \cite{KR}.

 The singular vectors $( v_{m_1}\ ,_{k} \ v_{m_2})$, $( (
 v_{m_1}\ ,_{k_1} \ v_{m_2}),_{k_2} \ v_{m_3})$,
 $(v_{m_1}\ ,_{k_1} \ (v_{m_2}\ ,_{k_2} \ v_{m_3}))$, and the matrix
 of $6j$-symbols can be visualized as follows.

\newpage

$$
 \tag1.2.5
 $$
   \vskip28ex
\ \newline \ \newline \ \newline \ \newline \ \newline \ \newline

 The braid group $B_3$ acts on the triple tensor
 products.  The basis
  $((v_{m_1}\ ,_{k_1} \ v_{m_2}),_{k_2} v_{m_3})$ is an
 eigenbasis for $R_{1,2}$, and the basis
 $(v_{m_1}\ ,_{k_1} \ (v_{m_2}\ ,_{k_2} \ v_{m_2}))$ is an eigenbasis
 for $R_{2,3}$.

 $$
 \tag1.2.6
 $$
 \vskip28ex
\ \newline \ \newline \ \newline \ \newline \ \newline \ \newline

 Consider the tensor product of $n$ irreducible modules.
 We describe its bases, the connecting matrices, and the
 $R$-matrix action.

 An {\it n-tree} is a planar tree with $n$ tops, one
 root, and $( n-1)$ internal triple vertices.  We
 numerate the tops by indices $1,\dots,n$ from left to
 right.

 See an example in (1.2.7):
 $$
 \tag1.2.7
 $$
\ \newline \ \newline \ \newline
\ \newline \ \newline \ \newline
\ \newline \ \newline \ \newline
 For an $n$-tree we denote the set of its internal
 vertices by In$_{T}$.  We say that an $n$-{\it tree has
 marked tops} if a nonnegative integer is assigned to
 every top.  Denote by $m_j$ the number assigned to the
 $j$-th top.

 For an $n$-tree $T$ with marked tops, we construct a
 basis in $L(m_1) \otimes\dots\otimes L(m_n)$, denoted
 by $\Cal B_T$.

 {\it A coloring} of $T$ is a map $c:\text{In}_T \to \Bbb
 Z_{\geq 0}$.  Define the {\it weight of a coloring} as the
 number
 $$
 m(c) = m_1+\dots+m_n -2\,
 \sum _{w \in {\text {In}}_T}  c(w).
 \tag1.2.8
 $$
 Delete an internal vertex $w$, then the tree is decomposed
 into three trees: {\it the left branch, the right branch}, and
 the part containing the root.

 Denote by $J^{\ell e} $ the set of indices of the tops lying
 in the left branch of $w$, denote by $I^ {\ell e}$ the set
 of the internal vertices lying in the left branch of $w$.
 For a coloring $c$, define the weight of the left branch
 as the number
 $$
 m^ {\ell e}(w)  = \sum_{j\in J^ {\ell e} }m_j - 2 \sum_{u\in
 I^ {\ell e} } c(u).
 $$
 Let $J^r,\ I^r$, and $m^r(w)$ be the sets and the weight
 defined for the right branch, analogously.

 Say that a coloring $c$ is {\it admissible at a vertex} $w$ if
 $c(w)\leq \text{ min }(m^ {\ell e}(w),\ m^r(w))$.  Say that
 a {\it coloring of an $n$-tree $T$ with marked tops is
 admissible} it is admissible at every internal vertex.

 The elements of the basis $\Cal B_T$ are numerated by
 the pairs $(c,\ell)$, where $c$ is an admissible coloring
 and $\ell =0,\dots, m(c)$.  The vector corresponding
 to $(c,\ell)$ is defined inductively on $n$.  Namely,
 let $w$ be the internal vertex of $T$ which is
 neighboring to the root of $T$.  Let $T^ {\ell e} $ and $T^r$
 be the left and right branches of $w$.  The coloring $c$
 induces colorings $c^ {\ell e} $ and $c^r$ of $T^ {\ell e} $ and
 $T^r$, respectively.  Let $1,\dots,i$ be the indices of
 the tops lying in the left branch of $w$.  Let $v^ {\ell e} $
 be the vector in $L(m_1) \otimes\dots\otimes L(m_i)$
 corresponding to the pair $( c^ {\ell e},0)$ and $v^r$ the
 vector in $L(m_{i+1}) \otimes\dots\otimes L(m_n)$
 corresponding to $( c^r,0)$.  If $i=1$, then $v^ {\ell e} =
 v_{m_1}$ where $v_{m_1}$ is the singular vector of
 $L(m_1)$.  If $i=n-1$, then $v^r=v_{m_n}$ where
 $v_{m_n}$ is the singular vector of $L(m_n)$.  Define
 the vector $v(c,\ell,T)$ by
 $$
 v(c,\ell,T) = f^{(\ell)} (v^ {\ell e},_{c(w)}v^r)
 \tag1.2.9
 $$
 where $( \cdot ,_{c(w)}\cdot)$ is defined by (1.2.1).

 The $U_q$-action is given by
\roster
\item"(1.2.10)" $ q^hv(c,\ell ,T) = q^{m(c)-2\ell} v(c,\ell ,T),$
\item"{}" $ev(c,\ell ,T) = (m(c)- \ell +1)_q v (c,\ell-1,T) \quad
 \text{for }\ell >0,$
\item"{}" $ ev(c,0,T) = 0,$
\item"{}" $fv(c,\ell ,T) = [\ell +1]_q v(c,\ell+1,T)
 \quad\text{for }\ell <m(c),$
\item"{}" $fv(c,m(c),T) = 0 .$
\endroster
  \vskip1ex

 Let $T$ and $T'$ be $n$-trees.  Let $w_1$ and $w_2$ be
 internal vertices of $T$ and $w'_1$ and $w'_2$ internal
 vertices of $T'$.  Say that $T$ and $T'$ are {\it
 adjacent} at $w_1,w_2, w'_1,w'_2$, if in a neighborhood
 of the vertices the trees have the form shown in
 (1.2.11), and the trees are identical outside the
 neighborhood.

\bigskip

 $$
 \tag1.2.11
 $$
 \vskip28ex

\ \newline \ \newline \ \newline
\ \newline \ \newline \ \newline
 Let $T$ and $T'$ be adjacent.  Let $c$ and $c'$ be
 admissible colorings of $T$ and $T'$ respectively.  Say
 that the {\it colorings are similar}, $c\equiv c'$, if the
 colorings are equal at the corresponding internal
 vertices of $T$ and $T'$ different from $w_1,w_2,
 w'_1,w'_2$, and if $c(w_1)+ c(w_2)= c'(w'_1)+ c'(w'_2)$.

 The matrix, connecting the bases $\Cal B_T$ and $\Cal
 B_{T'}$ of adjacent trees is given by
\bigskip
 $$
 v(c,\ell,T) =
 \sum_{c'\equiv c}\
 \biggl\{  \matrix m^ {\ell e} (w_1)\\c(w_1) \endmatrix
 \matrix m^r (w_1)\\c(w_2) \endmatrix
 \matrix m^r (w_2)\\c'(w'_1) \endmatrix \biggr\}_q\,
 v(c',\ell,T') \tag1.2.12
 $$
\bigskip
 where $\{ \ \}_q$ are $6j$-symbols.

 The matrix connecting the bases $\Cal B_T$ and $\Cal
 B_{T'}$ for arbitrary $T$ and $T'$ can be written as a
 product of the matrices connecting adjacent bases.

 For $j=1,\dots,n-1$, we describe the action of the
 operator
 $$
 R_{j,j+1} :
 L(m_1) \otimes\dots\otimes L(m_n) \to
 L(m_1) \otimes \dots L(m_{j+1}) \otimes
 L(m_j)\dots\otimes L(m_n).
 $$
 Denote the first tensor product by $L$ and the second by
 $L'$.

 We say that an $n$-tree $T$ is {\it adjacent to the
 transposition} of $j$ and $j+1$ if there exists an
 internal vertex $w$ such that the $j$-th top of $T$
 forms the left branch of $w$, and the $(j+1)$-th top of
 $T$ forms the right branch of $w$.  Let $T$ be adjacent
 to the transposition of $j$ and $j+1$.

 Let $\{ v(c,\ell,T;L) \}$ be the $T$-basis of $L$ and
 $\{ v(c,\ell,T;L') \}$ the $T$-basis of $L'$.  Then for
 every $( c,\ell )$ we have
 $$
 R_{j,j+1} v(c,\ell,T;L) =
 R(m_j,m_{j+1};c(w))\cdot v(c,\ell, T;L')
 \tag1.2.13
 $$
 where $R(m_j,m_{j+1};c(w))$ is defined by (1.2.3).  As
 an example see (1.2.3) and (1.2.6).

 Formulae (1.2.12) and (1.2.13) describe the action of
 the braid group $B_n$ on the tensor product of $n$
 irreducible $U_q$-modules.
 \vskip5ex
 \subhead (1.3) Crystal Base \endsubhead

 The tensor product $L(m_1)\otimes\dots\otimes L(m_n)$
 has the trivial basis $f^{(\ell_1)}v_{m_1}\otimes \dots
 \otimes f^{(\ell_n)}v_{m_n}$, $\ell_j=0,\dots, m_j,\
 j=1,\dots,n$.  It turns out that all the tree bases,
 constructed in (1.2), tend (in a appropriate sense)
 to the trivial basis as $q\to \infty$.  This phenomenon
 is described in \cite{K} and is called crystalization.

 For a finite dimensional $U_q$-module $M$ and
 $\lambda\in \Bbb Z$ let $M_\lambda= \{ u\in
 M\,\big\vert\, q^hu=q^{\lambda}u\}$  be the weight space
 of weight $\lambda$ .  We have $M= \underset \lambda\in \Bbb
 Z\to{\oplus}M_\lambda$.

 Any element $u$ of $M_\lambda$ is uniquely written in
 the form $u= \sum f^{(\ell)}u_\ell$ where $u_\ell
 \in\ker e\bigcap M_{\lambda+2\ell }$.  Following
 \cite{K} define the endomorphisms $\tilde{e}$ and
 $\tilde{f}$ on $M$ by
 $$
 \tilde{e} u=\sum f^{(n-1)}u_n  , \qquad
 \tilde{f} u= \sum f^{(n+1)}u_n.
 $$

 Let $A$ be the subring of $\Bbb Q(q^{1/4})$
 consisting of the rational functions of $q^{1/4}$
 regular at $q^{1/4}= \infty$.

 \vskip1ex

 {\it A crystal base} of $M$ is a pair $( L,B)$ satisfying the
 following conditions. \vskip1ex
 \roster
 \item"(1.3.1)" $L$ is a free sub-$A$-module of $M$ such
 that $M= \Bbb Q(q^{1/4})\otimes_{\negthickspace _A} L$.
 \endroster

 \roster
 \item"(1.3.2)" $B$ is a base of the $\Bbb Q $-vector
 space $L/q^{-1/4}L$.
 \endroster

 \roster
 \item"(1.3.3)" $L= \underset{\lambda\in\Bbb
 Z}\to{\oplus} L_\lambda$ and $B=
 \underset{\lambda\in\Bbb Z}\to{\amalg} B_\lambda$ where
 \endroster

 \vskip1ex
 \centerline{$\Lambda_\lambda = L\bigcap M_\lambda,\qquad
 B_\lambda = B\bigcap (L_\lambda /
 q^{-1/4}L_\lambda)$.}
 \vskip1ex

 \roster
 \item"(1.3.4)" $\tilde{e}L\subset L$ and
 $\tilde{f}L\subset L$.  Hence $\tilde{e}$ and
 $\tilde{f}$ operate on $L/q^{-1/4}L$.
 \endroster

 \roster
 \item"(1.3.5)" $\tilde{e}B\subset B\bigcup \{ 0\}$ and
 $\tilde{f}B\subset B\bigcup \{ 0\}$.
 \endroster

 \roster
 \item"(1.3.6)" For $b,\ b'\in B$, we have $b'=\tilde{f}b$
 if and only if $b=\tilde{e}b'$.
 \endroster

  \vskip1ex

 For $b\in B$, set $\epsilon(b)= \max \{ k\geq
 0\,\big\vert\, \tilde{e}^kb\neq 0\}$ and $\varphi(b)=
 \max\{ k\geq 0\,\big\vert\,\tilde{e}^kb\neq 0\}$.

 \vskip1ex

 \flushpar (1.3.7) {\bf Example.} For $m\in \Bbb Z_{\geq
 0}$, consider the irreducible module $L(m)$ with the
 basis $f^{(\ell)}v_m,\ \ell = 0,\dots, m$.  We have
 $\tilde{e}f^{(\ell)}v_m= f^{(\ell-1)}v_m$  and
 $\tilde{f} f^{(\ell)}v_m= f^{(\ell+1)}v_m$.  Let $\Cal
 L(m)= \underset{\ell}\to{\oplus} Af^{(\ell)}v_m$ and
 $\Cal B(m)= \{ f^{(\ell)}v_m\, \big\vert \, \ell =0,
 \dots, m\} \subset L/q^{-1/4}L$.  Then $(\Cal
 L(m),\ \Cal B(m))$ is a crystal base of $L(m)$.
 \vskip1ex
 \proclaim{(1.3.8) Theorem, \cite{K}}  Let $M_1$ and
 $M_2$ be two finite dimensional $U_q$-modules and let $(
 L_j,B_j)$ be a crystal base of $M_j$, $j=1,2$.  Set
 $L=L_1\otimes_{\negthickspace _A}L_2$ and $B=\{ b_1\otimes b_2\in
 L/q^{-1/4}L\, \big\vert \, b_j\in B_j\}$.
 \vskip.5ex

 \roster
 \item "(i)" Then $( L,B)$ is a crystal base of
 $M_1\otimes M_2$.\endroster

 \vskip1ex \roster
 \item "(ii)" For $b_j\in B_j,\ j=1,2,$ we have\endroster
 $$\align
 \tilde{f} (b_1\otimes b_2) &= \cases \tilde{f}
 b_1\otimes b_2 &\text{{\rm if }} \varphi (b_1) >\epsilon
 (b_2)\\
 b_1\otimes \tilde{f}b_2 &\text{{\rm if }}\varphi (b_1) \leq
 \epsilon (b_2),\endcases
 \vspace{2\jot}
 \tilde{e} (b_1\otimes b_2) &= \cases b_1\otimes
\tilde{e}b_2 &\text{{\rm if
 }}\varphi (b_1) < \epsilon (b_2)\\
 \tilde{e}b_1\otimes b_2 &\text{{\rm if }} \varphi (b_1) \geq
 \epsilon (b_2).\endcases
 \endalign$$
 \endproclaim

 \proclaim{(1.3.9) Theorem, \cite{K}}  Let $( L,B)$ be a
 crystal base of a finite dimensional
\newline $U_q$-module $M$.
 Then there exists an isomorphism $M\cong
 \underset{j}\to{\oplus} L(m_j)$ by which $( L,B)$ is
 isomorphic to $\underset{j}\to{\oplus}(\Cal L(m_j),\
 \Cal B(m_j))$.
 \endproclaim

  Consider the tensor product of $n$
 irreducible $U_q$-modules, $L= L(m_1)\otimes \dots
 \otimes L(m_n)$.  Let $\Cal L= \Cal L (m_1)\otimes_{\negthickspace _A}
\dots\otimes_{\negthickspace_A} \Cal L(m_n)$, $\Cal B=\Cal B
 (m_1)\otimes\dots\otimes \Cal B(m_n)$. By
 (1.3.8), $(\Cal L,\Cal B)$ is a crystal base of $L$.  We
 call $(\Cal L,\Cal B)$ the {\it distinguished crystal
 base} of the tensor product $L$.

 For any $n$-tree introduce
 $$\align
 &\Cal L_T = \bigoplus_{(c,\ell)} Av(c,\ell,T),\tag1.3.10\\
 & \Cal B_T = \{ v(c,\ell ,T)\} \subset \Cal
 L_T/q^{-1/4}\Cal L_T.
 \endalign$$

 It follows from (1.2.10) that $(\Cal L,\Cal B)$ is a
 crystal base of $L$.

 \proclaim{(1.3.11) Proposition}  For any $n$-tree $T$,
 we have $(\Cal L,\Cal B)=(\Cal L_T,\Cal B_T)$.
 \endproclaim

 \demo{Proof} It suffices to prove the proposition for
 $n=2$.  Let $(v_{m_1},_{k} \ v_{m_2})$ be the vector defined
 by (1.2.1).  Then for any $k=0,\dots,\min(m_1,m_2)$
 we have
 $$
 (v_{m_1}\ ,_{k}\ v_{m_2})=v_{m_1}\otimes f^{(k)} v_{m_2}
 \mod q^{-1/4}\Cal L.
 \tag1.3.12
 $$
 This equality easily implies that $\Cal L=\Cal L_T$ and
 then $\Cal B=\Cal B_T$.

 \proclaim{(1.3.13) Corollary}

For $k_1+k_2\leq m_2$,
 $$\align
 \biggl\{\matrix m_1&m_2&m_3\\ k_1&k_2&k\endmatrix
 \biggr\}_q &=
 \cases 1\text{ {\rm mod }}q^{-1/4}A &\text{{\rm if
 }} k=k_1\\
 0\text{ {\rm mod }}q^{-1/4}A &\text{{\rm
 otherwise.}}\endcases
 \intertext{For $k_1+k_2>m_2$,}
 \biggl\{\matrix m_1&m_2&m_3\\ k_1&k_2&k \endmatrix
 \biggr\}_q&= \cases 1\text{ {\rm mod }}q^{-1/4}A
 &\text{{\rm if }} k=m_2-k_1,\\
 0\text{ {\rm mod }}q^{-1/4}A &\text{{\rm
 otherwise.}} \endcases
 \endalign$$
 \endproclaim

 {\bf (1.3.14) Remark.}  For a finite dimensional
 $U_q$-module $M$, we have $\ker e =
 \underset{\lambda}\to{\oplus} \ker e \bigcap
 M_\lambda$.

 {\it A crystal base of $\ker e$ of} $M$ is a pair $(
 L_\lambda,B_\lambda)$ for each $\lambda$ such that
 \newline $\ker e\bigcap M_\lambda\neq \{ 0\}$.  A pair $(
 L_\lambda,B_\lambda)$ has to satisfy the following
 conditions.
 \roster
 \item"(1.3.15)"  $L_\lambda$ is a free sub-$A$-module of
 $\ker e\bigcap M_\lambda$ such that \newline $\ker e\bigcap
 M_\lambda= \Bbb Q (q^{1/4})\otimes
_AL_\lambda$.\endroster

 \vskip.5ex
 \roster
 \item"(1.3.16)"  $B$ is a base of the $\Bbb Q$-vector
 space $L_\lambda/q^{1/4}L_\lambda.$
 \endroster
 \vskip1ex

 Obviously, a crystal base of $\ker e$ is uniquely
 extended to a crystal base of $M$.  By (1.3.9), any
 crystal base of $M$ is the extension of a crystal base
 on $\ker e$.\enddemo

 \vskip4ex

 \subhead (1.4)  Bases in Tensor Products of
 $\bold{s}\pmb{\ell}_{\bold 2}$-Modules \endsubhead

 Consider the Lie algebra $\frak{g} = s\ell_2$ with the
 generators $e,f,h$ such that
$$
[e,f]=h ,  [ h,e ]=2e , [
 h,f ]=-2f .
$$
  Denote by $\Omega$ the Casimir element
 $$\frac{1}{2}h\otimes h+e\otimes f+f\otimes e\in
 \frak{g}\otimes \frak{g} .$$

 Let $M_1,\dots,M_n $ be $\frak{g}$ modules.  For $i<j$
 let $\Omega_{ij}$ be the linear operator on $M_1\otimes
 \dots\otimes M_n $ acting as $\Omega$ on $M_i \otimes
 M_i$ and as the identity operator on the other factors.

 For a $\frak{g}$-module $M$ and $\lambda\in\Bbb C$, let
 $M_\lambda=\{ u\in M\,\big\vert\, hu=\lambda u\}$ be
 the space of weight $\lambda$ and Sing$\,M_\lambda=\ker
 e\bigcap M_\lambda$ the space of singular elements of
 weight $\lambda$.

 For a nonnegative integer $m$, let $L(m)$ be the $(m+1)$
 dimensional irreducible $\frak{g}$-module.  The module
 is generated by its singular element $v_m$ such that
 $ev_m=0$ and $hv_m=mv_m$.  The elements $v_m,fv_m,\dots
 , f^mv_m$ form a basis of $L(m)$.

 The {\it Shapovalov form} on $L(m)$ is the bilinear form $B^m$
 such that
 $$
 B^m(f^kv_m,f^kv_m)= k!m!/(m-k)!, \qquad B^m(f^kv_m,f^\ell
 v_m)= 0
  \quad \text{for }k\neq \ell .
 $$

 The tensor product of irreducible $\frak{g}$ modules is
 the direct sum of irreducible modules: $L(m_1)\otimes
 L(m_n)= L(\vert m_1-m_2\vert)\,\otimes$ $L(\vert
 m_1-m_2 \vert+2)\otimes\dots\otimes L(m_1+m_2)$, and
 a singular vector of $L(m_1+m_2-2k) $ can be chosen of
 the form
 $$
 \{ v_{m_1}\ ,_k \ v_{m_2}\} =
 \sum^{k}_{p=0} (-1)^p\
 \frac{\overset{k-1}\to{\underset j=0\to{\prod}}(m_1+m_2
 -2k+j+2)}{\overset{p-1}\to{\underset
 j=0\to{\prod}}(m_1-j) \overset{k-p-1}\to{\underset
 j=0\to{\prod}}(m_2-j)}\
 f^pv_{m_1}\otimes f^{k-p}v_{m_2}.\tag1.4.1
 $$
 This decomposition gives a basis $\{ f^\ell \{
 v_{m_1}\ ,_k \ v_{m_2}\} \}$ in the tensor product.  We have

\roster
\item"(1.4.2)" $ \Omega f^\ell  \{ w_{m_1}\ ,_k \ v_{m_2}\} =\mu
 (m_1,m_2;k) \cdot
 f^\ell \{ w_{m_1} \ ,_k\ v_{m_2}\},$

\item"{}" $ \mu (m_1,m_2;k) =\frac{1}{2} m_1m_2 -
 k(m_1+m_2)+k(k-1).$
\endroster

 For nonnegative integers $m_1,\dots,m_n $, consider the
 tensor product $L=L(m_1)\otimes\dots\otimes L(m_n)$.
 For any $n$-tree $T$ the tensor product has a
 distinguished basis $\Cal B_T=\{ v(c,\ell ,T) \}$.  The
 elements of $\Cal B_T$ are numerated by the pairs $(
 c,\ell  )$ where $c$ is an admissible coloring of $T$
 and $\ell =0,\dots, m(c)$, see (1.3).  The vector
 $v(c,\ell ,T)$ is defined inductively on $n$ as in (1.3)
 but we use the formula
 $$
 v(c,\ell ,T) = f^\ell \{ v^ {\ell e} ,_{c(w)}v^r\}
 \tag1.4.3
 $$
 instead of formula (1.2.9).

 For any $n$-tree $T$ and a permutation $\sigma\in S_n$
 define a basis $\Cal B_{T,\sigma}=\{ v(c,\ell
 ,T,\sigma)\}$ in $L(m_1)\otimes\dots\otimes L(m_n)$.
 Namely, let $P_\sigma: L(m_1)\otimes\dots\otimes
 L(m_n)\to L(m_{\sigma(1)})\otimes\dots\otimes
 L(m_{\sigma(n)})$ be the permutation of factors.  Let
 $\Cal B_T =\{v(c,\ell ,T)\}$ be the basis in
 $L(m_{\sigma(1)}\otimes\dots\otimes L(m_{\sigma(n)})$
 corresponding to the tree $T$ with tops marked by
 $m_{\sigma(1)},\dots, m_{\sigma(n)}$. Set $v(c,\ell
 ,T,\sigma) = P^{-1}_\sigma v(c,\ell ,T)$ for all $c,\ell
 $.

 For any $T,\sigma$ and an internal vertex $w$ of
 $T$, define an operator $\Omega_{w,T,\sigma}$
 acting on $L(m_1)\otimes\dots\otimes L(m_n)$. Namely,
 for an internal vertex $w$, let $J$ be the set of
 indices of the tops of $T$ which belong to the right
 branch or the left branch of $w$.  Set
 $$
 \Omega_{w,T,\sigma} =
 \sum \Sb i<j\\i,j\in\sigma^{-1}(J)\endSb
 \Omega_{ij}.
 \tag1.4.4
 $$

 For every $T,\sigma,w$, the basis $\Cal
 B_{T,\sigma}$ is an eigenbasis for the operator
 $\Omega_{w,T,\sigma}$:
 $$
 \Omega_{w,T,\sigma}\  v(c,\ell ,T,\sigma) =
 \mu(c,T,\sigma,w)\cdot v(c,\ell ,T,\sigma)
 \tag1.4.5
 $$
 for all $c,\ell $ and suitable numbers
 $\mu(c,T,\sigma,w)$, cf. (1.4.2).

 \vskip4ex

 \subhead (1.5) The Selberg Integral and Normalizing
 Constants \endsubhead

 For any colored tree we define a function called the
 {\it normalizing constant.}

 The Selberg integral is the integral
 $$
 I_k(a,b;\kappa) = k!\int_{\Delta}
 \ \prod^k_{j=1}\ t_j^{-a/\kappa}
 (1-t_j)^{-b/\kappa}
 \prod_{1\leq i<j\leq k} (t_j-t_i)^{2/\kappa}
 dt_1\wedge\dots\wedge dt_n
 $$
 where $\Delta = \{t\in\Bbb R^k \,\big\vert\, 0<t_1<\dots
 <t_k<1\}$.  The integral can be computed explicitly
 \cite{M}:
 $$
 I_k(a,b;\kappa) =\ \prod^{k-1}_{j=0}\
 \frac{\Gamma (\frac{-a+j}{\kappa}+1)\Gamma
 (\frac{-b+j}{\kappa}+1)\Gamma (\frac{j+1}{\kappa}+1)}{\Gamma
 (\frac{-a-b+2k-j-2}{\kappa}+2)\Gamma (\frac1\kappa +1)}.
 \tag1.5.1
 $$

 Set
 $$
 q(\kappa) =\exp (2\pi i/\kappa).
 $$

 Define $J_k(a,b;\kappa)$ by
 $$\multline
 J_k(a,b;\kappa) = (2\pi \kappa)^{-k/2}
 \ q(\kappa)^{\frac{-kb}{2}+\frac{3}{4}k(k-1)}
 I_k(a,b,\kappa) \cdot \\ \vspace{1\jot}
\cdot \prod^{k-1}_{j=0} (1-\kappa/(a+b-2k+j+2))\cdot (b-j)_{q(\kappa)}
 \endmultline\tag1.5.2$$
 where $q$-numbers $(\ )_q$ are defined in (1.1.3).

 Let $T$ be an $n$-tree with tops marked by some
 nonnegative integers $m_1,\dots,m_n $.  Let $c$ be an
 admissible coloring of $c$.

 For an internal vertex $w\in{\text{In}}_{{}_T}$ define its {\it
 normalizing constant}  $C(w,c,T,\kappa)$ by
 $$
 C(w,c,T,\kappa) =
J_{c(w)}
 (m^ {\ell e} (w),  m^r(w);\kappa)
 $$
 where $c(w)$ is the color of $w$ with respect
 to $c$, and $m^ {\ell e} (w)$ (resp. $m^r(w)$) is
 the weight of the left (resp. right) branch of $w$,
 see (1.2).

 Define the {\it normalizing constant}, $C(c,T,\kappa)$, of a
 coloring $c$ by the rule
 $$
 C(c,T,\kappa) =
 \biggl( \sum_{w\in{\text{In}}_{{}_T}} c(w)\biggr) !
 \ \prod_{w\in{\text{In}}_{{}_T}} C(w,c,T,\kappa)/c(w)!\,.
 \tag1.5.3
 $$

\vskip4ex

 \head 2. Asymptotic Solutions to the
 Knizhnik-Zamolodchikov Equation \endhead

 \subhead (2.1)  Knizhnik-Zamolodchikov
 Equation\endsubhead

 Let $M_1,\dots,M_n $ be $\frak{g}=s\ell_2$ modules,
 $M=M_1\otimes \dots\otimes M_n $.  The
 Knizhnik-Zamolodchikov equation (KZ) on an $M$-valued
 function $\psi(z_1,\dots,z_n)$ is the system of
 equations
 $$\gather
 \kappa\,\frac{\partial\psi}{\partial z_i}=
  H_i\psi,\qquad i=1,\dots,n,\\
\intertext{where $k$ is a parameter of the equation and}
 H_i=\sum_{j\neq i} \frac{\Omega_{ij}}{z_i-z_j}.
 \endgather$$
 The KZ equation defines a connection on the trivial
 bundle $M\times\Bbb C^n\to\Bbb C^n$ with singularities
 at diagonal hyperplanes.  This connection is flat.
 Parallel translation with respect to this connection
 commutes with the $\frak{g}$ action on $M$.

 For a solution $\psi$ to the KZ equation we have
 $$(\frac{\partial}{\partial z_1}+\dots +
 \frac{\partial}{\partial z_n})\psi=0,
$$
 hence
 $\psi$ depends only on diferences $z_i-z_j$ of
 variables.

 For nonnegative integers $m_1,\dots,m_n $, let
 $L=L(m_1)\otimes\dots\otimes L(m_n)$. For any
 $\lambda$ the KZ equation with values in $L$ preserves
 Sing$\,L_\lambda$. The KZ equation with values in
 $\underset\lambda\to\oplus$ Sing$\,L_\lambda$
 determines the KZ equation on $L$.

 Denote by $B$ the bilinear form $B^{m_1}\otimes\dots
 \otimes B^{m_n}$ on $L$ , where $B^m$ is the Shapovalov
 form on $L(m)$. The operators $H_1,\dots,H_n$
 are $B$-symmetric: $$B(H_ix,y)=B(x,H_iy)$$ for all $i$
 and all $x,y\in L$ [RV].

 \vskip4ex

 \subhead (2.2)  Asymptotic Zones\endsubhead

 For $\sigma\in S_n$ let $$D_\sigma=\{z\in\Bbb
 R^n\,\vert\,z_{\sigma(1)} <\dots< z_{\sigma(n)}\}. $$
 For $\sigma\in S_n$ and an $n$-tree $T$, we define a
 diffeomorphism $$u_{T,\sigma}:D_\sigma\to\Bbb
 R\times (\Bbb R_{>0})^{n-1}$$ called an {\it asymptotic
 zone}.

 The first coordinate of $u_{T,\sigma}$ is $z_1+\dots
 +z_n$, the other coordinates are numerated by internal
 vertices of $T$.  Denote by $u_{\omega,T,\sigma}$ the
 coordinate corresponding to a vertex $\omega$.

 Let $\sigma=id$.  The coordinate $u_{\omega,T,id}$ is
 defined inductively.  For an internal vertex $\omega$,
 let $W$ be the set of internal vertices of $T$ belonging
 to the shortest path in $T$ from $\omega$  to the root.
 $W$ can be empty.  Assume that for any $y\in W$ the
 coordinate $u_{\omega,T,id}$  is already defined.  Let
 $J^ {\ell e}$ (resp. $J^r$) be the set of the tops of $T$
 belonging to the left (resp. right) branch of $w$.
 Set $j^ {\ell e}=\max \{j\in J^ {\ell e}\},\ j^r=\min (j\in
 J^r)$.
 $$
 u_{w,T,\text{id}}=
 (z_{j^r} - z_{j^ {\ell e}})\bigg/ \prod_{y\in W}
 u_{y,T,\text{id}}.
 \tag2.2.1
 $$

 For $\sigma\in S_n$, let $$\tau_\sigma: D_{\sigma}\to
 D_{\text{id}}, \qquad (z_1,\dots,z_n)\mapsto
 (z_{\sigma(1)},\dots,z_{\sigma(n)}).$$  Set
 $$u_{w,T,\sigma}=u_{w,T,\text{id}}\circ \tau_\sigma. $$

 Let $$\Cal U_n = \{ z_1,\dots,z_n\in\Bbb C^n \, \vert \,
 z_i\neq z_j \text{ for all }i,j\}.$$

 \proclaim{(2.2.2) Proposition}

 For every $\sigma,T$ the
 functions $z_1+\dots +z_n$, $\{ u_{w,T,\sigma} \}_{w\in
 {\text{In}}_{_T}}$ define a diffeomorphism $u_{T,\sigma}$ of
 $D_\sigma$ onto $\Bbb R\times (\Bbb R_{>0})^{n-1}$.
 The map $u_{T,\sigma}$ is extended to a biholomorphic map of
 $\Cal U_n$ to $\Bbb C\times (\Bbb C^\ast)^{n-1}$.
 Furthermore, $u^{-1}_{T,\sigma}$ is extended to a
 polynomial map of $\Bbb C^n$ to $\Bbb C^n$.
 \endproclaim

 Now we will construct curves connecting the domains
 $D_\sigma$ and lying in $\Cal U_n$.

 For $j\in \{ 1,\dots,n-1\}$ let $\tau $ be the
 transposition of $j$ and $j+1$.  Let $T$ be a tree
 adjacent to $\tau$, see (1.2), and $w$ the internal
 vertex of $T$ such that the $j$-th top of $T$ (resp. the
 $(j+1)$-th) forms the left (resp. right) branch of $w$.

 For any permutation $\sigma$, the
 biholomorphisms $u_{T,\sigma}$ and $u_{T,\sigma\tau}$
 have the property:
\roster
\item"{}"  $u_{w,T,\sigma}=-u_{w,T,\sigma\tau}$ ,
\item"{}"  $u_{y,T,\sigma}=u_{y,T,\sigma\tau}$ for every internal
 vertex $y$ different from $w$.
\endroster

 Define a curve $\gamma_{T,\sigma,\tau}:[0,1]\to
\Cal U_n$ by the rule:
\roster
\item"{}" $u_{w,T,\sigma}=\exp(\text{i}t)$ for
 $t\in [0,1]$,
\item"{}"  $u_{y,T,\sigma}=1$ for all  $t\in [0,1]$ and all $y\in
 \text{In}_T$ such that $y\neq w$.
\endroster
  We have $\gamma
 (0)\in D_{\sigma}$ and $\gamma (1)\in D_{\tau\sigma}$.
 The curve $\gamma_{T,\sigma,\tau}$ forms a half circle
 connecting $D_{\sigma}$ and $D_{\sigma\tau}$.  The
 complementary half circle is formed by the curve $\gamma
_{T,\sigma\tau ,\tau}$.

 For every $T,\sigma$ the biholomorphism $u_{T,\sigma}$
 resolves singularities of the union of hyperplanes in
 the following sense.  Let $$f:\Bbb C^n\to\Bbb C, \qquad
 (z_1,\dots,z_n) \mapsto \prod_{i<j}(z_i-z_j) .$$  Then
 for every $\sigma,T$ we have
 $$
  f\circ u_{T,\sigma}^{-1}=
  \prod_{w\in \text{In}_T}
 (u_{w,T,\sigma})^{a_w} \cdot
 (1+\Cal O(u))
 $$
 for suitable positive integers $\{ a_w \}$.

 Consider the KZ equation with values in $V=V_1\otimes
\dots\otimes V_n$.

 \proclaim{(2.2.3) Proposition}

For any $\sigma,T$, the
 KZ equation with respect to variables $z_1 + \dots + z_n$ and
 $\{ u_{w,T,\sigma}\}$ has the form
 $$\gather
 \biggl( \frac{\partial}{\partial z_1}+\dots +
 \frac{\partial}{\partial z_1} \biggr) \psi=0,\\  \vspace{3\jot}
 \frac{\partial \psi}{\partial u_{w,T,\sigma}}=\frac1\kappa
  \biggl( \frac{\Omega_{w,T,\sigma}}{u_{w,T,\sigma}} +
 \text{{\rm Reg}} \biggr) \psi ,
 \endgather$$
 where $\text{{\rm Reg}}$ is an {\rm End}$(V)-$valued
 function holomorphic at $u_{T,\sigma}=0$, and
 $\Omega_{w,T,\sigma}$ is the constant operator
defined in {\rm (1.4.4)}.
 \endproclaim

 {\bf Example.}  For the tree in (1.2.7) and the identity
 permutation, the coordinates are $z_1+z_2+z_3,\ u_1=
 z_3-z_2$, $u_2=( z_2-z_1)/(z_3-z_2)$, and the KZ
 equation has the form
 $$
 \frac{\partial\psi}{\partial u_1}=\frac1\kappa
 \biggl( \frac{\Omega_{12} + \Omega_{13}
 +\Omega_{23}}{u_1}
 + \text{Reg} \biggr), \qquad
 \frac{\partial\psi}{\partial u_2}=\frac1\kappa
 \biggl( \frac{\Omega_{12}}{u_2} + \text{Reg} \biggr).
 $$

 \vskip4ex

 \subhead (2.3) Asymptotic Solutions \endsubhead

 For $m_1,\dots,m_n\in\Bbb N$ and a nonnegative integer
 $\lambda$, consider the tensor product of irreducible
 $s\ell_2$ modules $L=L(m_1),\dots,L(m_n)$ and the KZ
 equation with values in Sing $L_\lambda$.  Throughout
 the remainder of the paper we assume that the parameter
 $\kappa$ of the KZ equation is not a rational number.

 For any $n$-tree $T$ and a permutation $w\in S_n$ we
 will construct a fundamental system of solutions,
 $\psi_{T,\sigma}$, to the KZ equation.

 Namely, consider the basis $\Cal B_{T,\sigma}$ of $L$.
 Then
$$\Cal B_{T,\sigma,\lambda} :=\Cal B_{T,\sigma}
 \,\cap {\text { Sing}}L_\lambda
$$ is a basis of
Sing$L_\lambda$.  Vectors of $B_{T,\sigma,\lambda}$ have the
 form $v(c,0,T,\sigma)$, see (2.1), where $c$ is an
 admissible coloring of the tree $T$ with tops marked by
 $m_{\sigma(1)},\dots,m_{\sigma(n)}$ and the coloring
 $c$ has weight $\lambda$, see (1.2.8).  Denote the set
 of such colorings by
\newline Adm $(T,\sigma,\lambda)$.  A vector
 $v(c,0,T,\sigma)$ is an eigenvector of the operators $\{
 \Omega_{w,T,\sigma}\}_{w\in{\text{In}}_{_T}}$ with eigenvalues
 denoted by $\{ \mu(c,T,\sigma,w) \}$, resp., see
 (1.4.5).

 For every such a vector there exists a unique solution,
 $\psi_{c,T,\sigma}$, to the KZ equation restricted to
 $D_{\sigma}$ such that
\roster
\item"(2.3.1)" $  \psi_{c,T,\sigma} = $
\item "{}" $ = C(c,T,\kappa) ( \prod_{w\in \text{In}T}
 (u_{w,T,\sigma})^{\mu(c,T,\sigma,w)/\kappa} ) \cdot
 (v(c,0,T,\sigma) + \Cal O(u_{T,\sigma}, \kappa) )$
\endroster
 where $C(c,T,\kappa)$ is the normalizing constant of the
 colored tree $T$ with tops marked by $m_{\sigma(1)}
,\dots,m_{\sigma(n)}$, $\Cal O(u_{T,\sigma}, \kappa)$ is a
  Sing$L_\lambda-$valued function which is regular
function of $u_{T, \sigma}$ at $u_{T,\sigma}=0$ and
which tends to
 zero as all coordinates $u_{T,\sigma}$ tend to zero,
  univalued branches of the functions $\{ u^\mu \}$ are
 chosen by the rule: arg $(u_{w,T,\sigma})=0$ for all
 $w\in {\text{In}}_{_T}$.

 The collection of these solutions form a fundamental
 system of solutions.  It will be called the {\it
 asymptotic solution corresponding to the asymptotic
 zone} $u_{T,\sigma}$, and will be denoted by
 $\psi_{T,\sigma}$.

 The asymptotic solution can be analytically continued to
 a system of multivalued solutions over $\Cal U_n$.

 \vskip4ex

 \subhead (2.4) Transition Functions Between Asymptotic
 Solutions \endsubhead

 The first main result of this paper describes transition
 functions between asymptotic solutions.

 To compare two asymptotic solutions $\psi_{T,\sigma}$
 and $\psi_{T',\sigma'}$ we have to distinguish a curve
 from $D_\sigma$ to $D'_\sigma$, lying in $ \Cal U_n$, then
 analytically continue $\psi_{T,\sigma}$ along the curve
 and express the analytic continuation in terms of $\psi
_{T',\sigma'}$.

 Let $T$ and $T'$ be $n$-trees.  Let $w_1$ and $w_2$ be
 internal vertices of $T$ and $w_1'$ and $w_2'$  internal
 vertices of $T'$.  Assume that $T$ and $T'$ are adjacent
 at $w_1,\ w_2\ w_1',\ w_2'$, see (1.2).

 For any $\sigma\in S_n$, the asymptotic solutions $\psi
_{T,\sigma}$ and $\psi_{T',\sigma}$ are defined over
 the same $D_\sigma$.

 \proclaim{(2.4.1) Theorem}

 For any $c\in\,\text{{\rm Adm}} \,(
 T,\sigma,\lambda)$ we have
 $$
 \psi_{c,T,\sigma}=
 \sum \Sb c'\equiv c\\c'\in\text{{\rm
 Adm}}\,(T',\sigma,\lambda)\endSb
 \biggl\{ \matrix m^ {\ell e} (w_1) &m^r(w_1) &m^r(w_2) \\
 c (w_1) &c(w_2) &c'(w_1')  \endmatrix \biggr\}_{q(\kappa)}
 \ \psi_{c',T',\sigma} .
 $$
 \endproclaim

 Here $\{ \  \}_{q(\kappa)}$ are the $6j$-symbols for
 $q(\kappa)=\exp (2\pi i/\kappa)$.  Definition of the arguments of
 the $6j$-symbols see in (1.2).

 The theorem is proved in section 4.

 For $j\in \{ 1, \dots,
 n-1\}$, let $\tau$ be the transposition of $j$ and
 $j+1$.  Let $T$ be an $n$-tree adjacent to $\tau$, see
 (1.2), and $w$ the internal vertex of $T$ such that the
 $j$-th tops of $T$ (resp. the $( j+1 )$) forms the left
 (resp. right) branch of $w$.

 For any $\sigma\in S_n$, the curve $\gamma
_{T,\sigma,\tau}$, defined in (2.2), connects
 $D_{\sigma}$ and $D_{\sigma\tau}$.  Continue $\psi
_{T,\sigma}$ along the curve and express the
 continuation in terms of $\psi_{T,\tau\sigma}$.

 \proclaim{(2.4.2) Proposition} For any $c\in\, \text{{\rm Adm}}
 \,(T,\sigma,\lambda)$ we have
 $$
  \psi_{c,T,\sigma}=
 R(m_{\sigma(j)},m_{\sigma(j+1)};c(w))_{q(\kappa)}
  \psi_{c,T,\sigma\tau}
 $$
 where $R$ is defined in (1.2.3).
 \endproclaim

 The proposition easily follows from the definition of
 the normalizing constant.

 Statements (2.4.1) and (2.4.2) allow us to compare two
 arbitrary asymptotic solutions.

 By (2.4.1) and (2.4.2), the transition functions between
 asymptotic solutions to the KZ equation are exactly the
 same as the transition functions between the bases in tensor products of
 irreducible modules over the quantum group
 $U_{q=q(\kappa)}$, where $U_{q=q(\kappa)}$ is the
 $\Bbb C$-algebra obtained from $U_q$ by specializing the
 indeterminate $q^{1/4}$ to $q(\kappa/4)$.  Namely, for
 natural $m_1,\dots,m_n $, let $L(m_1), \dots, L(m_n)$ be
 the irreducible $s\ell_2$-modules with highest weights
 $m_1,\dots,m_n$, resp.  For a nonrational $\kappa\in\Bbb C$
 and an integer $\lambda$, consider the KZ equation with
 the parameter $\kappa$ and the values in Sing $L_\lambda$.
 For an $n$-tree $T$ and a permutation $\sigma \in S_n$,
 consider the asymptotic solution $\psi_{T,\sigma}(\kappa)=
 \{ \psi_{c,T,\sigma}\}$.

 Let $L(m_1,q=q(\kappa)),\dots,L(m_n,q=q(\kappa))$ be irreducible
 $U_{q=q(\kappa)}$ modules with highest weight
 $m_1,\dots,m_n$, resp.  For a permutation $\sigma\in
 S_n$, consider the tensor product
 $$\gather
 L^{\sigma}=
 L(m_{\sigma(1)}, q=q(\kappa)) \otimes \dots\otimes
 L(m_{\sigma(n)},q=q(\kappa)) \tag2.4.3\\
\intertext{and the subspace}
 \text{Sing } L^{\sigma} (q=q(\kappa))=
 \{ v\in L^\sigma\,\vert\, q^hv=q(\kappa)^\lambda v,\ ev=0\}.\tag2.4.4
 \endgather$$
 Let $\Cal B_{T,\sigma}(q=q(\kappa))=\{ v(c,0,T)\}$ be the
 basis in Sing $L^\sigma(q=q(\kappa))_\lambda$ corresponding
 to an $n$-tree $T$, see (1.2).

 Define a map
 $$
 \pi_{T,\sigma}(\kappa):\Cal B_{T,\sigma} (q=q(\kappa))
 \to \psi_{T,\sigma}(\kappa), \qquad v(c,0,T) \mapsto
 \psi_{c,T,\sigma}.\tag2.4.5
 $$

 \proclaim{(2.4.6) Corollary of (1.2.12) and (2.4.1)}

 For
 every $\sigma\in S_n$ the maps $\{ \pi_{T,\sigma}(\kappa)\}$
 induce a well defined isomorphism $\pi_\sigma(\kappa)$ of {\rm
 Sing} $L^\sigma(q=q(\kappa))_\lambda$ and the space of
 solutions to the KZ equation over $D_\sigma$ with parameter
$\kappa$ and values in Sing$L_\lambda$ .
 \endproclaim

 \proclaim{(2.4.7) Corollary of (1.2.13) and (2.4.2)}

Under the isomorphisms described in {\rm (2.4.6)}, the
 $R$-matrix action on
\newline $\{ \text{{\rm Sing} }
 L^\sigma(q=q(\kappa))_\lambda\}_{\sigma\in S_n}$ given by
 {\rm (1.2.13)} is isomorphic to the monodromy of the KZ
 equation with the parameter $\kappa$ and with values in {\rm
 Sing} $L_\lambda$.
 \endproclaim

 Kohno \cite{Ko} and Drinfeld \cite{D} proved
 existence of an isomorphism between the $R$-matrix
 action and the monodromy representation of the KZ
 equation for generic $\kappa$.  The case $\kappa=1/\ell $ for a
 natural $\ell$ is described in \cite{V, \S\S 13--14}.  In
 Section 3 we'll construct  isomorphisms (2.4.6) and
 (2.4.7) geometrically in terms of integral
 representations for solutions to the KZ equation, see
 [SV, V].
\vskip3ex

 \subhead (2.5)  Quasiclassical Asymptotics \endsubhead

 Let $D $ be a ball, $\pi:\Bbb C^N\otimes D\to D$
 projection.  Let
 $$
 \nabla_\kappa = \kappa d - \omega
 $$
 be a holomorphic connection in $\pi$ depending on the
 parameter $\kappa$.  Here
 $$
 \omega = H_1 dz_1+\dots +H_n dz_n
 $$
 where $\{ H_i\}$ are matrix valued functions.  Assume
 that for every $\kappa$ the connection is integrable,
 $\kappa d\omega+\omega\wedge \omega=0$, or
 $$
 \frac{\partial H_i}{\partial z_j} =
 \frac{\partial H_j }{\partial z_i} ,\qquad [H_i,H_j]=0
 \tag2.5.1
 $$
 for all $i$ and $j$.

 An {\it asymptotically flat section} is a section of the
 form
 $$
 F=\exp (S/\kappa) (f_0+\kappa f_1+\dots \,)\qquad \text{for }\kappa \to
 0.
 \tag2.5.2
 $$
 Here $S(z_1,\dots,z_n)$ is a function, $\{
 f_j(z_1,\dots,z_n)\}$ are sections of $\pi$, and $F$ must
 be a formal solution of the equation $\nabla_\kappa F=0$.

 We  call $\exp(S/\kappa)f_0$ an {\it asymptotically flat
 section of the first order} if there exists a power
 series (2.5.2) which provides an asymptotic solution
 to the equation $\nabla_\kappa F=0$ modulo terms of order
 $\kappa^2$.

 Assume that the linear operators $\{ H_j\}$ are
 simultaneously diagonalizable for each $z\in D $: there
 exists a basis $\{ v_\ell (z)\}$ in $\Bbb C^N$ such
 that
 $$
 H_j(z)v_\ell (z) =
  \lambda_{j\ell } (z)\cdot v_\ell (z)
 $$
 for all $j,\ \ell $.  Assume that the spectrum of $\{
 H_j\}$ separates elements of the basis:  for every $z$
 and every $\ell ,\ m$, there exists $j$ such that
 $\lambda_{j\ell}(z)\neq \lambda_{jm}(z)$.

 Let $F= \exp(S/\kappa)f_0$ be an asymptotically flat section of
 the first order, then $f_0$ is an eigenvector of the
 operators $\{ H_i\}$,
 $$
 H_j(z) f_0(z)=\lambda_{j} (z) f_0(z),\qquad j=1,\dots
 ,n,
 \tag2.5.4
 $$
 moreover, $\lambda_j= \frac{\partial S}{\partial z_j}$,
 see \cite{RV}.

 Let $\exp(S/\kappa)f_0$ and $\exp(T/\kappa)g_0$ be asymptotically
 flat sections of the first order corresponding to the
 same eigenvector of $\{ H_i\}$, then
 $$
 \exp (S/\kappa)f_0=\,\text{const}\,\cdot \exp(T/\kappa)g_0,
 \tag2.5.5
 $$
 see \cite{RV}.

 Assume that there exists a symmetric bilinear form
 $B:\Bbb C^n\otimes \Bbb C^n\to \Bbb C$ and the operators
 $H_1,\dots, H_n$ are symmetric with respect to $B:
 B(H_ix,y)=B(x,H_iy)$ for all $x,y,i$.  Assume that
 $\exp (S/\kappa)(f_0+\dots \,)$ and $\exp (T/\kappa)(g_0+\dots
 \,)$ are two asymptotically flat sections.

 \proclaim{(2.5.6) Lemma [RV]}  If $S-T\neq\text{{\rm const}}$
,  then $B(f_0,g_0)\equiv 0$, if $S-T=\text{{\rm
 const}}$ then $B(f_0,g_0)\equiv\text{{\rm const}}$.
 \endproclaim

 The KZ equation gives an example of a family of flat
 connections.  The KZ operators $\{ H_i\}$ are symmetric
 with respect to the Shapovalov form.

 For natural $m_1,\dots,m_n $, let $L(m_1),\dots,L(m_n)$
 be the irreducible $s\ell_2$ modules with highest
 weights $m_1,\dots,m_n $.  For a nonrational number $\kappa$
 and an integer $\lambda$, consider the KZ equation with
 parameter $\kappa$ and with values in Sing $L_\lambda$.  For
 an $n$-tree $T$ and a permutation $\sigma\in S_n$
 consider the asymptotic solution $\psi_{T,\sigma}= \{
 \psi_{c,T,\sigma}\}$.  Consider the coordinates $\{
 u_{w,T,\sigma}\}_{w\in {\text{In}}_{_{T}}}$. These coordinates
 take positive values on $D_\sigma$.

 For $\epsilon>0$, let
 $$
 D _{\sigma,T,\epsilon}=
 \{ p\in D_\sigma\,\vert\,u_{w, T,\sigma}(p)<\epsilon
 \text{ for all }w\in{\text{In}}_{_{T}}\}.
 $$
 \vskip3ex

 Our second main result is the following theorem.
 \proclaim{(2.5.7) Theorem}

 Assume that $\kappa=is$ and $s\to
 +0$.  Then there exists $\epsilon>0$ such that any
 solution $\psi_{c,T,\sigma}\in \psi_{T,\sigma}$,
 restricted to $D_{\sigma,T,\epsilon}$ , has an asymptotic
 expansion
 $$
 \psi_{c,T,\sigma} \sim \prod_{w\in {\text{In}}_{_{T}}}
 (u_{w,T,\sigma})^{-i\mu(c,T,\sigma,w)/s}
 \exp(-iS/s) \sum^{\infty}_{j=0} f_js^j
 \tag2.5.8
 $$
 where $\{ \mu\}$ are the numbers described in {\rm (2.3.1)}
  and {\rm (1.4.5)};
 $$
S=S_{c,T,\sigma}:D_{\sigma,T,\epsilon}\to\Bbb R, \qquad
 f_j=f_{j,c,T,\sigma}:D_{\sigma,T,\epsilon}\to {\text{\rm
 Sing}}L_\lambda
$$
are suitable real analytic functions,
 all the functions $\{S,f_j\}$ depend only on $\{
 u_{w,T,\sigma}\}_{w\in{\text{In}}_{_{T}}}$ and can be
 analytically continued to real analytic functions in a
 neighborhood of the set $\{ u_{w,T,\sigma}=0
 \,\vert\,w\in{\text{In}}_{_{T}}\}$.

 The asymptotic expansion can be differentiated an
 arbitrary number of times.
\endproclaim

 The asymptotic expansion means that for any $N$
 $$
 \bigg\vert \psi_{c,T,\sigma} \prod _{w\in {\text{In}}_T}
 (u_w)^{i\mu(w)/s} \exp (iS/s) - \sum^N_{j=0} f_js^j
 \bigg\vert=\Cal O (s^{N+1})
 $$
 uniformly in $D_{\sigma,T,\epsilon}$.

 \vskip2ex

 The theorem has the following  appendix.

 \proclaim{(2.5.9) Appendix}
\roster
\item  For any $p\in
 D_{\sigma,T,\epsilon}$ the first terms $\{
 f_{0,c,T,\sigma}(p)\}$ of the above asymptotic
 expansions form a basis in {\rm Sing}$L_\lambda$
 orthogonal with respect to the Shapovalov form.

\item For any $c$, $ f_{0,c,T,\sigma}(p)$ \ \ tends to
\item"{}"
$$(-1)^k k!\cdot
 \biggl( \prod_{w\in{\text{In}}_{_{T}}} (c(w)!)^3\,
 \prod^{c(w)-1}_{j=0}
 \frac{(m^ {\ell e} (w)+m^r(w) - 2c(w)+j+2)^3}{(m^ {\ell e}
 (w)-j)(m^r(w)-j)}\biggr)^{\negmedspace^{-1/2}}
 \cdot v(c,0,T,\sigma)
 $$
 as $u_{w,T,\sigma}(p) \to 0$ for all $w\in
{\text{In}}_{_{T}}$.  Here  $T$ is an $n$-tree with vertices
 marked by $m_{\sigma(1)},\dots, m_{\sigma(n)}$.  The
 numbers $c(w),\ m^ {\ell e}(w),$ and $m^r(w)$ for an
 $n$-tree with marked tops are defined in (1.2).  The
 vector $v(c,0,T,\sigma)$ is defined by (2.3.1).
\endroster
 \endproclaim

The Theorem and the Appendix are proved in Section 4.

 \vskip4ex

 \subhead (2.6) Quasiclassical Asymptotics and Crystal
 Base \endsubhead

 For natural $m_1,\dots,m_n $, let
 $L(m_1,q),\dots,L(m_n,q)$ be irreducible $U_q$ modules
 with highest weights $m_1,\dots,m_n$, resp.  For any
 permutation $\sigma\in S_n$, let
 $$
 L^\sigma(q) =
 L(m_{\sigma(1)},q)\otimes\dots\otimes
 L(m_{\sigma(n)},q) .
 $$
 Statements (2.4.6) and (2.5.7) allow us to give a
 construction of a crystal base in $L^\sigma(q)$ purely
 in terms of quasiclassical asymptotics of solutions to
 the KZ equation.

 This \lq\lq quasiclassical" crystal base coincides with
 the distinguished crystal base of the tensor product
defined in (1.3).  This statement can be considered as a
 quasiclassical characterization of the distinguished
 crystal base.

 We will give a construction for $\sigma=id$, since for
 an arbitrary permutation the construction is the same.

 To construct a crystal base in $L(q)= L(m_1,q) \otimes
 \dots\otimes L(m_n,q)$ it suffices to construct a
 crystal base in
 $$
 \text{Sing } L(q)_\lambda =
 \{ v\in L(q)\,\vert\, q^hv =q^\lambda v,\ ev=0\}
 $$
for every $\lambda$.

 Let $\mu :\Bbb Q (q^{^{\frac14}})\to\frac14\Bbb Z$ be the
 valuation map which assigns to a function
 $g(q^{^{\frac14}})$ the order of its pole at
 $q=\infty$.

 Using the KZ equation, we'll construct a map
$$
\mu
 :\text{Sing} \ L(q)_\lambda\to\frac14\Bbb Z
$$
such that
 $$\align
& \mu (v) = 0 \quad\text{ iff } v=0,
 \tag2.6.1\\ \vspace{1\jot}
&\mu (v+w)  \leq \max (\mu (v),\mu (w)),\\  \vspace{1\jot}
& \mu (g v) = \mu (g)+\mu (v)
 \endalign$$
 for all $v,w\in$ Sing $L(q)_\lambda,\ g\in \Bbb Q
 (q^{^{\frac14}})$.

 Let $v\in $ Sing $L(q)_\lambda$,
 $$
  v= \sum_{k_1,\dots,k_n} a_K f^{k_1} v_1\otimes\dots
 \otimes f^{k_n}v_n,
 $$
 where $v_j$ is the generating vector of $L(m_j,q)$,
 $j=1,\dots,n$, and the coefficients $\{ a_K\}$ are
 rational functions in $q^{^{\frac14}}$.

 For $\kappa\in\Bbb C$, let Sing $L(q=q(\kappa))_\lambda$ be the
 space defined by (2.4.4).  For any $\kappa$ such that
 $q=q(\kappa)$ is not a pole of the functions $\{ a_K\}$, let
 $v(q=q(\kappa))\in$ Sing $L(q=q(\kappa))$ be the specialization of
 $v$ at $q=q(\kappa)$.

 Consider the KZ equation with parameter $\kappa$ and values
 in Sing $L_\lambda$, see (2.3).  Let Sol $(\kappa)_\lambda$
 be the space of its solutions over $D_{{\text{id}}}$, and let
 $\pi(\kappa):$ Sing $L(q=q(\kappa))\to$ Sol $(\kappa)_\lambda$ be the
 isomorphism defined in (2.4.6).  Then
 $y(\kappa)=\pi(\kappa)(v(q=q(\kappa)))\in$ Sol $(\kappa)_\lambda$ is a
 solution over $D_{{\text{id}}}$ to the KZ equation with parameter
 $\kappa$.

 For any $n$-tree $T$, let $\psi_{T,\text{id}}(\kappa)= \{
 \psi_{c,T,\text{id}}\}$ be the asymptotic solution, then
 $$
 y(\kappa)=\sum_c b_c(q(\kappa)^{^{\frac14}})\cdot \psi_{c,T,\sigma}
 $$
 where $\{b_c(q^{^{\frac14}})\}$ are some rational
 functions.

 \proclaim{(2.6.2) Proposition}

\flushpar {\bf{\smc 1.}} Let $\Vert \
 \Vert:\text{{\rm Sing }}L_\lambda\to\Bbb R_{\geq 0}$
 be any norm on {\rm Sing} $L_\lambda$.  Let $\kappa=is,\ s\in
 \Bbb R$, and $s\to +0$.  For any $n$-tree $T$, let
 $D_{\text{{\rm id}},T,\epsilon}\subset D_{\text{{\rm
 id}}}$ be the domain described in {\rm (2.5.7)}.  Then for
 every $p\in D_{\text{{\rm id}},T,\epsilon}$ there exists
 a limit of $s\cdot \ln\Vert y(is)(p)\Vert$.  This limit is an
 element of $\frac14\Bbb Z$, and does not depend on $p$
 and $T$.  Denote this limit by $\mu (v)$.

 \flushpar {\bf{\smc 2.}}  The map $\mu$, defined by this rule, has
 properties {\rm (2.8.1)}.  \endproclaim

 \demo{Proof} The proposition easily follows from
 (2.5.7), (2.5.9), (2.4.1), and (1.3.12).
 \enddemo

 Let $A\subset \Bbb Q (q^{^{\frac14}})$ be the subalgebra
 of functions regular at $q=\infty$.  Let
 $$
 \Cal L^0_\lambda=
 \{ v\in \text{Sing}\,L(q)_\lambda\,\vert\, \mu (v)\leq
 0\}.
 $$

 \proclaim{(2.6.3) Corollary} $\Cal L^0_\lambda$ is an
 $A-$submodule of $\thinspace\text{{\rm Sing}}\thinspace L(q)_\lambda$.
 \endproclaim

 Let $(\Cal L,\Cal B)$ be the distinguished crystal base
 in $L(q)$ defined in (1.3).  Let $(\Cal L_\lambda,\Cal B_\lambda)$ be the
 crystal base in $\text{Sing}\,L(q)_\lambda$ induced by
 $(\Cal L,\Cal B)$.

 \proclaim{(2.6.4) Proposition} $\Cal L^0_\lambda= \Cal
 L_\lambda$.
 \endproclaim

 The Proposition easily follows from (2.5.7) and (2.5.9).

 For any $n$-tree $T$, let $\psi_{T,\text{{\rm id}}}= \{
 \psi_{c,T,\text{{\rm id}}}\}$ be the asymptotic solution
 considered in (2.5.7) and (2.5.9), let
 $$
 \biggl\{ \prod_w (u_{w,T,\text{{\rm id}}})^{-\mu
 (c,T,\text{{\rm id}},w)/s}
  \exp(-iS_{c,T,\text{{\rm id}}}/s) \cdot f_{0,c,T,\text{{\rm
 id}}}\biggr\}
 $$
 be the collection of its first terms of asymptotics.

 Denote by Asym$_T$ the $\Bbb Q-$module of
$\Bbb Q-$linear combinations of the first terms of asymptotics.

 \proclaim{(2.6.5) Proposition} The above construction
 induces the canonical isomorphism of the $\Bbb Q$
 modules $\Cal L^0_\lambda/ q^{^{-\frac14}}\Cal
 L^0_\lambda$ and {\rm Asym}$_{_T}$.
 \endproclaim

 The module Asym$_{_T}$ has a canonical basis generated by
 the first terms of asymptotics.  This basis is uniquely
 determined by the fact that each element of the basis is
 the first term of an asymptotic expansion of the form
 (2.5.8) of a solution to the KZ equation and by the
 normalizing condition (2.5.9.2).  The canonical basis in
 Asym$_{_T}$ induces a basis in $\Cal L^0_\lambda/
 q^{^{-\frac14}}\Cal L^0_\lambda$.  This induced basis
 coincides with the basis $\Cal B_\lambda$.

 \vskip4ex

\head 3. Integral Representations for Solutions to the
 KZ Equation and the Bethe Vectors \endhead

 \subhead (3.1)  Local System \endsubhead

 Let $m_1,\dots,m_n\in\Bbb C,\ \kappa\in\Bbb C^\ast,$ let $ k$ be a
 nonnegative integer.  Set
 $$
 \Phi(t,z)=\prod_{1\leq \ell <j\leq n}(z_j - z_\ell
)^{^{m_\ell m_j/2\kappa}}
 \prod_{1\leq \ell <j \leq k}(t_\ell -t_j)^{^{2/\kappa}}\cdot
 \prod^k_{\ell =1}\prod^n_{j=1}(z_j - t_\ell
)^{^{-m_j/\kappa}}. \tag3.1.1
 $$
 $\Phi$ is a multivalued holomorphic function on
 $$
 \Cal U_{k+n}=
 \{ (t,z)\in \Bbb C^{k+n}\,\vert\, t_\ell \neq t_m,\
 t_\ell \neq z_m,\ z_\ell \neq z_m,\ \text{for all }\ell
 ,m\}.
 $$

 Univalued branches of $\Phi$ over open subsets of
 $\Cal U_{k+n}$ generate a complex one-dimensional local
 system over $\Cal U_{k+n}$ denoted by $\Cal S(\kappa)$.

 Let $ \Cal U_{n}=\{ z\in \Bbb C^n\,\vert\,z_\ell\neq z_j\
 \text{for all }\ell,j\},$ and let
$$
 pr_{k,n}:\Cal U_{k+n}\to\Cal U_{n}, \qquad (t,z) \mapsto
 z,
 $$
 be a projection.  Denote its fiber $pr^{-1}(z)$ by
 $\Cal U_{k,n}(z)$.  Denote by $\Cal H_{k,n}(\kappa)$ the
 complex vector bundle over $\Cal U_{n}$ with fiber
 $H_k(\Cal U_{k,n}(z),\Cal S(\kappa))$ over $z\in \Cal U_n$.  The
 bundle has a canonical flat connection called the
 {\it Gauss--Manin connection}.

 The symmetric group $S_k$ acts on $\Cal U_{k+n}$ by
 permutations of coordinates $t_1,\dots,t_k$ preserving
 fibers of the projection .  The function $\Phi$ is
 symmetric with respect to this action.  Therefore, $S_k$
 naturally acts on the singular chains in $\Cal U_{k,n}(z)$
 with coefficients in $\Cal S(\kappa)$.  This action induces
 an action of $S_k$ on $H_k(\Cal U_{k,n}(z),\Cal S(\kappa))$.
 This action on fibers of $\Cal H_{k,n}(\kappa)$ commutes
 with the Gauss-Manin connection.

 Denote by $H_k(\Cal U_{k,n}(z),\Cal S(\kappa))_{-}$ the skew
 symmetric part of this action :
 $$
 H_k ( \Cal U_{k,n}(z),\Cal S(\kappa))_- =
 \{ v\in H_k\,\vert\,\sigma v=(-1)^{\negthinspace^{\vert\sigma\vert}}v
 \text{ for all }\sigma\in S_k\}.
 $$
 \vskip4ex

\subhead (3.2) Integral representations \endsubhead

 For $m_1,\dots,m_n \in \Bbb N$ and a nonnegative integer
 $\lambda$, consider the tensor product of irreducible
 $s\ell_2$ modules, $L=L(m_1)\otimes \dots \otimes
 L(m_n)$, and the KZ equation with parameter $\kappa$ and
 values in Sing$\,L_\lambda$.  We will describe integral
 representations for solutions to the KZ equation.

 Set $k=(m_1+\dots+m_n -\lambda)/2$. The number  $k$ is a
 nonnegative integer. {\it A monomial of weight} $\lambda$ is
 an element of $L_\lambda$ of the form
 $$
 f_K = f^{k_1} v_1\otimes \dots \otimes f^{k_n} v_n
 \tag3.2.1
 $$
 where $K=(k_1,\dots ,k_n)$, $k_1+\dots +k_n=k$.  For a
 monomial $f_K$ define a differential $k$-form in $t$ and
 $z$:
 $$\gather
 \eta(f_K) = A_K(t,z) dt_1\wedge \dots\wedge
 dt_k,\tag3.2.2 \\
 \vspace{3\jot}
 A_k = \sum_{\sigma\in S(k;k_1,\dots ,k_n)} \
 \prod^k_{i=1} \ \frac{1}{(t_i-z_{\sigma(i)})} .
 \endgather$$
 The sum is over the set $S(k;k_1,\dots ,k_n)$ of maps
 $\sigma$ from $\{ 1, \dots , k\}$ to $\{ 1, \dots , n
 \}$ such that for all $m$ the cardinality of
 $\sigma^{-1}(m)$ is $k_m$.

 Consider the $L_\lambda$-valued form
 $$
 N=  \sum_{f_K\in L_\lambda}\Phi(t,z) \cdot \eta(f_K) \otimes
 f_K
 \tag3.2.3
 $$
 where $\Phi$ is the function defined by (3.1.1).  $N$ is
 a multi-valued holomorphic $k$-form on $\Cal U
 _{k+n}$.

 In \cite{SV} it is proved that

\roster
\item"{(3.2.4)}"  For every $i$, the form
 $$
 \biggl(\kappa \frac{\partial }{\partial z_j} - \sum_{\ell \neq j}
 \frac{\Omega_{j,\ell }}{z_j-z_\ell}\biggr) N
 $$
 is a sum of differential of a suitable $( k-1)$-form
 and a form which has zero restriction to fibers of
 the projection $pr_{k,n}$.

 \vskip1ex

 \item"{(3.2.5)}"  The form
 $eN=\sum \Phi \cdot\eta(f_K)\otimes ef_K$ is a sum of
 differential of a suitable $( k-1)$-form and a form
 which has zero restriction to fibers of $pr_{k,n}$.

 \vskip1ex

 \item"{ (3.2.6)}" The forms mentioned in
 (3.2.4) and (3.2.5) have the shape $\sum\Phi w(M)\otimes
 M$ where the sum is over monomials in $L$, and $\{ w(M)
 \}$ are suitable rational forms, regular on
 $\Cal U_{k+n}$.
\endroster

 Assume that $\gamma (z)\in H_k(\Cal U_{k,n}(z),\Cal
 S(\kappa))$.  Assume that the map $z\mapsto \gamma (z)$ forms
 a flat section of the bundle $\Cal H_{k,n}(\kappa)$ when $z$
 runs through an open subset of $\Cal U_n$.  The the
 function
 $$
 \Psi (z) = \int_{\gamma (z)}N
 \tag3.2.7
 $$
 takes values in Sing$\,L_\lambda$ and satisfies the KZ
 equation with parameter $\kappa$ [SV, V].

 \vskip1ex

 \flushpar {\bf (3.2.8)  Remark.}  The group $S_k$
 naturally acts on the space of differential forms on
 $\Cal U_{k+n}$ by permutations of $t_1,\dots,t_k$.  The
 differential form $N$ is skew symmetric with respect to
 this action.  Let $\nu : H_k(\Cal U_{k,n}(z),\Cal S(\kappa)) \to
 H_k(\Cal U_{k,n}(z),\Cal S(\kappa))_-$ be the canonical
 projection, then for any flat section $z\mapsto \gamma
 (z)$ of $\Cal H_{k,n}(\kappa)$ we have
 $$
 \int_{\gamma (z)} N = \int_{\nu (\gamma (z))} N.
 $$
 \vskip4ex
 \subhead (3.3) Quasiclassical Asymptotic Solutions to
 the KZ Equation \endsubhead

 Let $\kappa\to 0$.  We use the form $N$ defined in (3.2.3) to
 construct quasiclassical asymptotic solutions to the KZ
 equation.

 The function $\Phi$ can be written in the form
 $$\gather
 \Phi (t,z) = \exp (S(t,z)/\kappa)\\
 \intertext{where}
 S(t,z) = \sum_{1 \leq \ell <j\leq n} \ \frac{m_\ell
 m_j}{2}
 \ \ln (z_j - z_\ell ) +
 \sum_{1 \leq \ell <j\leq n} 2\ln (t_\ell -t_j) -
 \sum^{k}_{\ell =1} \sum^{n}_{j=1} m_j\ln (z_j - t_\ell ).
 \tag3.3.1\endgather$$
 Set
 $$
 \text{Hess}_t(-S) = \det \biggl( -\frac{\partial ^2S}{\partial
 t_\ell\partial t_j} \biggr).
 \tag3.3.2
 $$

 For a fixed $z\in \Cal U_n$, consider the equation of
 critical points of $S$ in $\Cal U_{k,n}(z)$:
 $$
 \frac{\partial S}{\partial t_j}=0,\qquad j=1,\dots ,k .
 \tag3.3.3
 $$

 Let $t=t(z)$ be a nondegenerate solution of (3.3.3)
 holomorphically depending on $z$ in a neighborhood of a
 point $z^0\in \Cal U_n$.

 Let $B(z)\subset \Cal U_{k,n}(z)$ be a small ball with
 center at $(t(z),z)$.  Set
 $$
 B_-(z) = \{ (t,z)\in \Cal U_{k,n}(z)\  \big\vert \ \text{Im}
 (S(t,z))
 < \text{Im} (S(t(z),z))\}.\tag3.3.4
 $$

 It is known that $H_k(B(z),B_-(z),\Cal S(\kappa))$ is
 one-dimensional, see \S11 in \cite{AGV}.  The Gauss--
 Manin connection identifies these groups for neighboring $z$'s.

 A generator of the homology group ( for all $\kappa$
 simultaneously ) can be chosen as follows.  Fix a branch
 of arguments of all functions $t_j-t_\ell ,\ t_j-z_\ell
 ,\ z_\ell -z_m$ for all $\ell ,m$ in a neighborhood of
 the point $(t(z^0),z)$.  This choice determines a
 branch, $\beta (\kappa)$, of $\Phi$ and a branch of $S$ in a
 neighborhood of $(t(z^0),z^0)$.  The branch $\beta (\kappa)$
 gives a section of $\Cal S(\kappa)$ for all $\kappa$.

 There exist local coordinates $u_1,\dots ,u_k$ in
 $\Cal U_{k,n}(z)$ centered at $t(z)$ such that the
 coordinates holomorphically depend on $z$ in a
 neighborhood of $z^0$ and
 $$
 S(t(u),z) =
 -i(u^2_1+\dots +u^2_n) +g(z)
 $$
 for some function $g(z)$.  For these coordinates and a
 small $\epsilon>0$, denote by $d(z)$ the disc $\{
 (u,z)\in \Cal U_{k,n}(z)\, \vert \, u_1,\dots ,u_k \in
 \Bbb R,\ u^2_1+\dots +u^2_k\leq \epsilon\}$.

 The homology class of the cycle
 $$
 \delta(z,\kappa) = (d(z),\beta (\kappa)\vert_{_{d(z)}})
 \tag3.3.5
 $$
 gives a generator of $H_k(B(z),B_-(z),\Cal S(\kappa))$.  The
 classes $[\delta(z,\kappa)]$ form a flat section of the
 Gauss-Manin connection of the vector bundle over a
 neighborhood of $z^0$ with fiber $H_k(B(z),B_-(z),\Cal
 S(\kappa))$.

 Set
 $$
 \Psi(z,\kappa) = (2\pi \kappa)^{^{-\frac{k}2}}\ \int_{\delta(z,\kappa)}\  N.
 \tag3.3.6
 $$

 \proclaim{(3.3.7) Theorem [RV]} Let $\kappa=is,\ s\in \Bbb
 R$, and $s\to +0$.
 \flushpar {\bf 1.}  Then the function $\Psi(z)$ has
 an asymptotic expansion
 $$
 \Psi (z,is) \sim \exp (-iS(t(z),z)/s)
 \sum^{\infty}_{j=0} g_j (z) s^j.
 $$
 where $\{ g_j \}$ are {\rm Sing}$L_\lambda-$valued
 holomorphic functions defined in a neighborhood
 of $z^0$.
 \flushpar {\bf 2.} The function $\Psi(z,\kappa)$ gives a
 quasiclassical asymptotic solution to the KZ equation in
 the sense of (2.5).
 $$
 g_0(z) = \text{{\rm
 Hess}}_t(-S(t(z),z))^{^{-\frac12}}
 \sum_{f_K\in L_\lambda} A_K (t(z),z)f_K. \tag"\text{{\bf
 3.}}"
 $$
 \endproclaim

 \vskip1ex

 \flushpar {\bf (3.3.8) Remark.} Let $\delta^0(z,\kappa)$
 be a singular chain in $B(z)$ with coefficients in $\Cal
 S(\kappa)$ such that the boundary of $\delta'(z,\kappa)$ lies in
 $B_-(z)$ and the class of $\delta'(z,\kappa)$ in
 $H_k(B(z),B_-(z),\Cal S(\kappa))$ coincides with
 $[\delta(z,\kappa)]$.  Then the function
 $$
 \Psi^0(z,\kappa) = (2\pi \kappa)^{^{-\frac{k}2}}
 \int_{\delta^0(z,\kappa)}\ N
 $$
 has the same asymptotic expansion as the function
 $\Psi(z,\kappa)$.

 The vector
 $$
 g(t(z),z) = \sum_{f_K\in L_\lambda} A_K (t(z),z)f_k
 \tag3.3.9
 $$
 is called the {\it Bethe vector}, see [B,Ba,BF,FFR,G,R,TV].

 It is shown in \cite{V2}, that
 $$
 B(g(t(z),z),g(t(z),z)) = \text{{\rm Hess}}_t(S(t(z),z))
 \tag3.3.10
 $$
 where $B$ is the Shapovalov form on Sing$\,L_\lambda$.

 The group of permutations of coordinates $t_1,\dots,t_k$
 acts on the set of $t$-critical points of $S$.  The
 Bethe vectors corresponding to $t$-critical points of
 the same orbit are identical.  If two critical points
 $t=t^1(z)$ and $t=t^2(z)$ lie in different orbits, then
 the corresponding Bethe vectors are
 orthogonal with respect to the Shapovalov form,
 $$
 B(g(t^1(z),z),g(t^2(z),z)) = 0.
 \tag3.3.11
 $$
 Moreover, for generic $z\in \Cal U_n$, there are exactly
 $\dim \text{Sing}\,L_\lambda$ different orbits of
 nondegenerate $t$-critical points, and, consequently, the
 corresponding Bethe vectors form a basis in
 Sing$\,L_\lambda$, see \cite{RV}.
\vskip3ex

 \subhead (3.4) Homology Class Sitting on a Critical
 Point
 \endsubhead

 Under the assumptions of Section (3.3), assume that for
 almost all $\kappa$ a homology class  $[\gamma (\kappa)]\in
 H_k(\Cal U_{k,n}(z^0),\Cal S(\kappa))$ is given.  Assume that a
 chain
 $$
 \gamma (\kappa) = \sum^M_{j=1} (c_j,\alpha_j(\kappa))
 \tag3.4.1
 $$
 is given, such that:
 \roster
 \item "{(3.4.2)}" For every $j$, $c_j\subset
 \Cal U_{k,n}(z^0)$ is a singular cell.

 \item "{ (3.4.3)}" For every $j$, $\alpha_j(\kappa)$ is a
 section of $\Cal S(\kappa)$ over $c_j$ which has the following form:
 \item "${}$"  Fix a branch of arguments of all functions
 $t_j-t_\ell,\ t_j-z_\ell,\ z_\ell-z_m$ for all $\ell
 ,m$ in a neighborhood of $c_j$.  This choice determines
 a branch $\beta _j(\kappa)$ of $\Phi$ over $c_j$.  Then
 $\alpha_j(\kappa)= \epsilon_j(\kappa)\beta _j(\kappa)$ where
 $\epsilon_j(\kappa)$ is a rational function of
 $q(\kappa)^{\frac14}$.

 \item "{ (3.4.4)}"  If $\kappa \in \Bbb C$ is such that
 $\epsilon_1(\kappa), \dots , \epsilon_m(\kappa)$ are defined, then the
 chain $\gamma (\kappa)$ is a cycle representing $[\gamma
 (\kappa)]$.
 \endroster

In this case we will say that {\it the class $[\gamma (\kappa) ]$
is flat with respect to $\kappa$, and $\gamma (\kappa)$ is a flat chain
representative.}

  Assume that $[\gamma (\kappa)]$ is flat.  Assume
 that $[\gamma (\kappa)]$ has a representing chain $\gamma
 (\kappa)$ of the form (3.4.1)--(3.4.4) and such that

 \roster
\item"{(3.4.5)}"  $(c_1,\alpha_1(\kappa))$ has the form
 $(d(z^0),\epsilon_1(\kappa)\beta (\kappa)\big\vert_{d(z^0)})$
 where $(d(z^0),\beta
 (\kappa)\big\vert_{d(z^0)})$  is described in (3.3.5) and
 $\epsilon_1(\kappa)$ is a rational function of
 $q(\kappa)^{\frac14}$.

 \item"{ (3.4.6)}"  For $\kappa=is$, $s\in \Bbb R_{_{>0}}$, and
 for every $j=2, \dots , M$, we have
 $$
 \lim_{s\to +0}\
 \sup_{c_j}
 \vert \alpha_j(\kappa) \vert\, \big/ \,
 \vert \epsilon_1(\kappa)\beta(\kappa)(t(z^0),z^0) \vert  =0.
 $$
\endroster

 In this case we will say that {\it the class $[\gamma (\kappa)]$
 sits on the $t$-critical point} $(t(z^0),z^0)$.

 Assume that $[\gamma (\kappa)]$ sits on $(t(z^0),z^0)$.  Let
 $[\gamma (\kappa)]\in H_k(\Cal U_{k,n}(z),\Cal S(\kappa))$ be the
 class such that for a fixed $\kappa$ the map $z\mapsto
 [\gamma (\kappa)]$ forms a flat section of the Gauss--Manin
 connection over a neighborhood of $z^0$ and $[\gamma
 (z^0,\kappa)]= [\gamma (\kappa)]$.

 \proclaim{(3.4.7) Theorem}  Under the above conditions
 consider the {\rm Sing}$\,L_\lambda$-valued function
 $$
 \Psi(z,\kappa) = \int_{[\gamma (z,\kappa)]} N
  $$
 which for a fixed $\kappa$ gives a solution to the KZ
 equation with parameter $\kappa$.  Assume that $\kappa=is,\ s\in
 \Bbb R$, and $s\to +0$.  Then

 \flushpar {\bf {\rm 1.}}  The function $\Psi$ has an
 asymptotic expansion
 $$
 \Psi(z,is) \sim F(\kappa) \exp (-iS(t(z),z)/s)
 \sum^{\infty}_{j=0} g_j(z)s^j
 $$
 where $\{ g_j \}$ are {\rm
 Sing}$\,L_\lambda$-valued holomorphic functions defined
 in a neighborhood of $z^0$.
 \flushpar {\bf {\rm 2.}} The function $\Psi$ gives a
 quasiclassical asymptotic solution to the KZ equation in
 the sense of (2.5).
 \flushpar {\bf {\rm 3.}} $g_0(z)$ is given by {\rm (3.3.7.3)}.
 \endproclaim

 The theorem is a corollary of (3.3.7).

 Consider the natural action of $S_k$ on singular chains
 and homology classes.  Let $[\gamma (z,\kappa)]$ be as above.
 Define a new skew symmetric class $[\gamma (z,\kappa)]_- \in
 H_k(\Cal U_{k,n}(z),\Cal S(\kappa))_-$ by
 $$
 [\gamma (z,\kappa)]_-=\sum_{\sigma\in S_k}
 (-1)^{^{\vert\sigma\vert}}
 \sigma [\gamma(z,\kappa)].\tag3.4.8
 $$

 We have
 $$
 \underset{[\gamma (z,\kappa)]_-}\to{\int\,N}
 = k! \underset{[\gamma (z,\kappa)]}\to{\int\,N}.\tag3.4.9
 $$

 \proclaim{(3.4.10) Corollary} The integral in (3.4.9)
 has an asymptotic expansion as $\kappa=is,\ s\in \Bbb R,$ and
  $s\to +0$.  The asymptotic expansion is equal to the
 asymptotic expansion in (3.4.7.1) multiplied by $k!$.
 \endproclaim

 We will say that $[\gamma (z,\kappa)]_-$ {\it sits on the $S_k$
 orbit of the $t$-critical point $(t(z),z)$.}

 \vskip4ex

 \subhead (3.5) Example of a Bethe Vector \endsubhead

 Let
 $$\gather
 \Phi_{k,m_1,m_2,\kappa} (t ) =
 \prod^k_{j=1}(- t_j)^{-\frac{m_1}{\kappa}}
 (1-t_j )^{-\frac{m_2}{\kappa}}\ \cdot
\prod_{1\leq j<\ell\leq k} (t_j-t_\ell)^{2/\kappa} ,
 \tag3.5.1\\ \vspace{4\jot}
 S(t) = S_{k,m_1,m_2} (t,\kappa) =\kappa \ln\Phi (t ).
 \endgather$$

 \proclaim{(3.5.2) Theorem, [V2], cf. [Sz, \S6.7]}

 \flushpar {\bf 1.}  If $t^0=(t_1,\dots,t_k)$ is a
 critical point of $\Phi$ then the symmetric functions
$$
\lambda_1= t_1+\dots+t_k , \qquad \lambda_2= \sum t_it_j, \qquad
 \dots , \qquad \lambda_k= t_1\cdot \dots \cdot t_k
$$
are given by
 $$
 \lambda_\ell  =
 \pmatrix k\\ \ell\endpmatrix \ \prod^{\ell }_{j=1}\
 \frac{(m_1+j-k)}{(m_1+m_2+j+1-2k)}
 $$
 for all $\ell$. \vskip1ex

 \flushpar {\bf 2.}  The Bethe vector in {\rm
 Sing}$\,(L(m_1)\otimes L(m_2))_{m_1+m_2-2k}$
 corresponding to the critical point $t^0$ is given by
 $$\gather
 \sum^k_{p=0} (-1)^p \pmatrix k\\ p \endpmatrix \
 \frac{\prod^{k-1}_{j=0}(m_1+m_2-2k+j+2)}{\prod^{p-1}_{j=0
 } (m_1-j) \prod^{k-p-1}_{j=0} (m_2-j)}
 \ f^p\,v_1\otimes f^{k-p}\,v_2. \\
 \vspace{3\jot} \text{{\rm Hess}}\,(S(t^0)) =
 k!\, \prod^{k-1}_{j=0}
 \frac{(m_1+m_2-2k+j+2)^3}{(m_1-j)(m_2-j)}.\tag"{\bf
 3.}"\endgather
 $$
 \endproclaim

\bigskip

 \subhead (3.6) Asymptotics of Critical Points and Bethe
 Vectors\endsubhead
\bigskip

 In (2.2) we have constructed asymptotic zones.  In this
 section we will describe asymptotics of $t$-critical
 points and Bethe vectors in an asymptotic zone, cf.
 \cite{RV}.

 Asymptotic zones are numerated by $n$-trees and elements
 of the symmetric group  $S_n$.  We will consider the
 case of the identity permutation.  The case of an
 arbitrary permutation is treated similarly.

 For an $n$-tree $T$ consider the asymptotic zone
 $$
 u_{T,id} = \{ z_1+\dots +z_n ,\{ u_{w,T,id}\}
 _{w\in {\text{In}}_T}\}
 : D_{id} \longrightarrow \Bbb R\times (\Bbb
 R_{>0})^{n-1} . \tag3.6.1
 $$

 Consider the function $\Phi(t,z)$ defined by (3.1.1).

 Mark the tops of $T$ by $m_1,\dots,m_n$.  Let the set {\rm Adm} be the
 set of all admissible colorings of $T$ having weight
 $\lambda= m_1+\dots+m_n-2k$, see (1.2).

 \proclaim{(3.6.2) Theorem}

  Assume that  $z\in D_{id}$ and $u_{w,T,id}(z)<<1$ for
 all $w\in {\text{In}}_T$.  Then we have the following three
 statements.

 \flushpar {\bf 1.}  The number of $S_k$ orbits of
 nondegenerate $t$-critical points of $\Phi(t,z)$ is
 equal to $\dim \text{{\rm Sing}}L_\lambda$ , see {\rm \cite{RV}}.

 \flushpar {\bf 2.}  The $S_k$ orbits of nondegenerate
 $t$-critical points can be numerated by elements of {\rm
 Adm} in such a way that the $S_k$-orbit corresponding to
 a coloring $c\in \text{{\rm Adm}}$ has a $t-$critical point
 $(t^c(z),z)$ of the following form:

 Let $\{ M_w\}_{w\in {\text{In}}_T}$ be an arbitrary
 partition of $\{ 1,\dots ,k \}$ into a union of disjoint
 subsets such that $\#M_w=c(w)$. For each $M_w$, fix its
 arbitrary ordering $M_w= \{ j_1(w),\dots
 ,j_{c(w)}(w)\}$. For any $w\in {\text {In}}_T$ define a number $l(w)$
as follows. The tops of the tree $T$ are numerated by 1,...,$n$ from
left to right. The number $\ell(w)$ is the maximum of the indices of the tops
lying on the left branch of the tree $T$ at $w$. Let
$m^ {\ell e}(w)$ and $m^r(w)$
be the numbers defined in (1.2) for a vertex $w$ of a colored tree $T$
with marked tops. Consider the function  $ S_{c(w),m^ {\ell e}(w),m^r(w)}$
defined by (3.5.1). Let $(a_1(w),\dots
 ,a_{c(w)}(w))$ be coordinates of a critical point of
 the function
 $ S_{c(w),m^ {\ell e}(w),m^r(w)}$. Then the critical point $(t^c(z),z)$
has the form:
 $$
 t^c_{j_p(w)} (z) - z_{\ell (w)} =
 (a_p(w) + \Cal O (u_{T,id}(z)))
  \prod_y \,u_{y,T,id}
 $$
 The product is over all internal vertices $y$ lying on the
 shortest path in $T$ connecting $w$ and the root of
 $T$ (the vertex $w$ is included ) .
 The function $\Cal O$ is a function of $u_{x,T,id}$,
 $x\in {\text{In}}_T$, holomorphic in a neighborhood of the set
 $\{ u_{x,T,id}=0,\ x\in {\text{In}}_T \}$, and such that $\Cal O
 (u=0)=0$.

 \flushpar {\bf 3.}  The Bethe vector corresponding to
 $(t^c(z),z)$ has the following form:
 $$
 g(t^c(z),z) = (v(c,0,T,id) +\Cal O (u(z)))
 \prod_{w\in {\text{In}}_T} (
 u_{_{w,T,id}})^{-c(w)-b(w)}
 $$
 where $b(w)= \sum_y c(w)$, the sum is over $y\in
 {\text{In}}_{T}$ such that $y$ lies on the right or on the left
 branch of $\,T$ at $w$.  The function $\Cal O(u)$ has the
 same structure as in {\rm (3.6.2.2)}.
 \endproclaim

 To prove the theorem we consider  new variables $\{
 x_{j_p(w)},u_{T,id}\}$ defined by
 $$\gather
 t_{j_p}(w) -z_{\ell (w)} \ = \
 x_{j_p}(w) \prod_y u_{y,T,id},\tag3.6.3\\
 z_m= z_m(u_{T,id}),
 \endgather$$
 where the product is the same as in (3.6.2.2), and the
 second formula is determined by (3.6.1).  It is easy to
 see that
 $$\multline
 \Phi(t(x,u),z(u)) = \\
 \vspace{2\jot} (-1)^{A/\kappa}
 \prod_{w\in {\text{In}}_T} (u_{w,T,id})^
 {\mu (c,T,id,w)/\kappa} \cdot
 \biggl( \prod_{w\in {\text{In}}_T} \Phi_{c(w),m^\ell
 (w),m^r(w),\kappa}
 (x(w)) +\Cal O (u) \biggr)
 \endmultline\tag3.6.4
 $$
 where $A$ is some constant, $\{ \mu (c,T,id,w) \}$ are
 the eigenvalues of the vector $v(c,0,T,id)$ with respect to
 the operators $\{ \Omega_{w,T,id} \}$, see (2.3.1),
 $\Cal O$ is a function of $x,u$ holomorphic in a
 neighborhood of the set
 $$
 X = \{ x_{j_p} (w) = a_p(w),u_{w,T,id}=0\quad\text{for
 all }w,p\}
 \tag3.6.5
 $$
 and which is zero on $X$.  This statement implies
 (3.6.2.2) and (3.6.2.3).  Statement (3.6.2.1) is proved in
 \cite{RV}.

 \vskip1ex
 \flushpar (3.6.6) {\bf Remark.}  There are integral
 representations for solutions to the KZ equation with
 values in a tensor product of modules over an arbitrary
 Kac-Moody Lie algebra $\frak{g}$ , see [SV, V1].  There are
 obvious $\frak{g}$ analogs of (3.3.7), (3.4.4) and
 (3.6.2), cf \cite{RV}.

 \head 4. Integral Representations for Solutions to the KZ
 Equation and \newline \ \ the Quantum Group $U_q$ \endhead
\bigskip
 \flushpar {\bf (4.1) Complex} \cite{V1, \S5}.

 For $\kappa \in \Bbb C$, consider the $\Bbb C$-algebra
 $U_{q= q(\kappa)}$.  For $k>0$, let
$$(U^-_{q=
 q(\kappa)})_k = \Bbb C f^k\subset  U_{q= q(\kappa)}\qquad {\text{
 and}} \qquad U^-_{q= q(\kappa)} = \underset
 k>0\to{\oplus}\Bbb C f^k .
$$
The map $\mu :U ^-\to
 U ^-\otimes U ^-$ ,
 $$
 \mu  : f^k \mapsto \sum^{k-1}_{\ell =1}
 {k \choose l}_{q(\kappa)}
 f^\ell \otimes f^{k-\ell}
 \tag4.1.1
 $$
 defines a coalgebra structure on $ U ^-$, here
 $$
{k\choose l}_{q(\kappa)} =
 {(k)_{q(\kappa)}!\over (\ell)_{q(\kappa)}!
 (k-\ell)_{q(\kappa)}!}
$$
 is the $q$-binomial coefficient.

 For a $ U _{q=q(\kappa)}$-module $M$ with highest
 weights, the map $\nu:M\to  U^- _{q=q(\kappa)}\otimes
 M$,
 $$
 \nu : u\mapsto \sum_{\ell>0} f^\ell  \otimes
 \frac{(q^{-\frac{h}{4}}e)^\ell }{[\ell ]_{q(\kappa)}!}\,
 u,
 \tag4.1.2
 $$
 defines a $U^- _{q=q(\kappa)}$-comodule structure on
 $M$.

 For $k\geq 0$, set
$$C^k(M)= ( U^-
 _{q=q(\kappa)})^{\otimes k}\otimes M .
$$
  Define $d:
 C^k\to C^{k+1}$ by
 $$\multline
 d: a_k\otimes \dots \otimes a_1\otimes u
 \mapsto -a_k\otimes \dots \otimes a_1\otimes \nu(u)
 \\+ \sum^k_{\ell =1}(-1)^{\ell +1} a_k \otimes \dots
 \otimes \mu (a_\ell ) \otimes \dots \otimes a_1 \otimes
 u.\endmultline
 \tag4.1.3
 $$

 For any $\lambda$ and $k$, set
 $$
 C^k_\lambda(M) = \underset{\ell _1,\dots ,\ell
 _k}\to{\oplus}
 ( U ^-)_{\ell _1} \otimes \dots \otimes ( U
 ^-)_{\ell _k}
 \otimes M_{\lambda-\ell _1-\dots -\ell _k}.\tag4.1.4
 $$
 Then $C^k= \underset \lambda\to{\oplus}\,
 C^k_\lambda$.  The differential $d$ preserves the
 grading.  For any $\lambda$ we have a complex
 $(C^\bullet _\lambda (\ M),d)$ and
 $$
 H^0 (C^\bullet _\lambda ,d) = \ker e\cap M_\lambda
 \tag4.1.5
 $$
 for a nonrational $\kappa$.  Assume that $\kappa$ is
 not a rational number.

 For an integer $m$, let $V(m,q=q(\kappa))$ (resp.
 $L(m,q=q(\kappa))$ be the $ U _{q=q(\kappa)}$ Verma
 module (resp. irreducible module) with highest weight
 $m$ .

 For any $\sigma\in S_n,\ \lambda\in \Bbb Z$, and natural
 $m_1,\dots,m_n $, set
 $$\gather
 V^{\sigma}(q=q(\kappa)) = V(m_{\sigma(1)},q=q(\kappa))
 \otimes \dots \otimes V(m_{\sigma(n)},q=q(\kappa)) ,
 \tag4.1.6\\
 \text{Sing}\, V^{\sigma}(q=q(\kappa))_\lambda =
 (V^{\sigma})_\lambda\cap \ker e.
 \endgather$$
 Let $L^{\sigma}(q=q(\kappa))\ \text{and Sing}\,
 L^{\sigma}(q=q(\kappa))_\lambda$ be the corresponding
 objects constructed from the irreducible modules.
 Denote by
 $$
 pr : \text{Sing}\, V^{\sigma}(q=q(\kappa))_\lambda
 \to \text{Sing}\, L^{\sigma}(q=q(\kappa))_\lambda
 \tag4.1.7
 $$
 the natural epimorphism.

 For any $j=1,\dots ,n-1$, we have an isomorphism
 $$
 R_{j,j+1} : ( U ^-)^{\otimes k} \otimes
 V^{\sigma} \to  (U ^-)^{\otimes k} \otimes
 V^{\sigma\tau}
 \tag4.1.8
 $$
 where $\tau \in S_n$ is the transposition of $j$ and
 $(j+1)$.  $R_{j,j+1}$ commutes with the differential and
 induces an isomorphism of cohomology groups of the
 corresponding complexes.
 \vskip4ex

 \subhead (4.2)  The Main Result\endsubhead

 Assume that $\kappa\in \Bbb C$  is not a rational number.
 Consider the function $\Phi(t,z)$ defined by (3.1.1) and
 the objects associated with the function $\Phi(t,z)$ in
 (3.1) and (3.2).

 For $z\in \Cal U _n$, let $\Cal C _{\bullet }(\Cal U
 _{k,n} (z),\Cal S(\kappa))$ be the complex of singular
 chains in $\Cal U _{k,n} (z)$ with coefficients in $\Cal
 S(\kappa)$.  The group $S_k$ of permutations of
 coordinates $t_1,\dots,t_k$ acts on $\Cal C_{\bullet}$.  Denote by
 $\Cal C_{\bullet }(\Cal U _{k,n}(z),\Cal S(\kappa))_-$
 the skew symmetric part of the action.

 \proclaim{(4.2.1) Theorem}  For any $\sigma\in S_n$ and
 any $z\in D_\sigma$ there exists a monomorphism of
 complexes
 $$
 \nu(z,\kappa) : C^{\bullet }_\lambda
 (V^{\sigma}(q=q(\kappa)))
  \to \Cal C _{k-\bullet } (\Cal U _{k,n}(z),\Cal
 S(\kappa))_-,
$$
where $ \lambda=m_1+\dots+m_n -2k.
 $
 Denote the image of $\nu$ by
$$C_\bullet (\Cal U_{k,n}(z),\Cal S(\kappa))_- \subset
\Cal C _{\bullet } (\Cal U _{k,n}(z),\Cal
 S(\kappa))_- .
$$
The monomorphism has the
 following six properties.

 \vskip1ex
 \flushpar {\bf {\rm 1.}}  $\nu(z,\kappa)$ is a
 quasiisomorphism. Denote by the same symbol
$\nu_{\*}(z,\kappa)$ the induced isomorphism of
homology groups:
$ H^{\bullet}( C^{\bullet }_\lambda
 (V^{\sigma}(q=q(\kappa))),d)\cong
 H_{k-\bullet}(\Cal U _{k,n}(z),\Cal S(\kappa))_-$.
 \vskip1ex

 \flushpar {\bf {\rm 2.}}  $\nu_{\*}(z,\kappa)$ is flat
 with respect to the Gauss-Manin connection.  Namely, if
 $$P(z^1,z^2): H_\bullet (\Cal U _{k,n}(z^1),\Cal
 S(\kappa))_- \to H_\bullet (\Cal U _{k,n}(z^2),\Cal
 S(x))_-$$ is the isomorphism of the Gauss-Manin connection along a curve
in $D_\sigma$ from a point $z^1 \in D_\sigma$ to a point $z^2 \in D_\sigma$,
then $\nu_{\*}(z^2,\kappa)=
 P(z^1,z^2)\cdot\nu_{\*}(z^1,\kappa)$. \vskip1ex

 \flushpar {\bf {\rm 3.}}  Let $\tau$ be the transposition
 of $j$ and $j+1$ for $j=1,\dots ,n-1$.  Connect the
 domains $D_\sigma$ and $D_{\sigma\tau }$ by a curve
 $\gamma :[0,1]\to \Cal U _n$ of the following form.  Let
 $(z^0_1,\dots ,z^0_n)\in  D_\sigma,\ z^0_{\sigma(1)}
 <\dots < z^0_{\sigma(n)}$.  Assume that $z_{\sigma(j)}$ and
 $z_{\sigma(j+1)}$ are very close.  Set $z_\ell (\gamma
 (s)) = z^0_{\ell }$ for $\ell \neq \sigma(j+1)$, set
 $z_{\sigma(j+1)}(\gamma (s)) = \exp (\pi\,is)(
 z^0_{\sigma(j+1)} -z^0_{\sigma(j)}) + z^0_{\sigma(j)}$.
 Let
$$
P_\gamma :
 H_{\bullet }(\Cal U _{k,n} (\gamma (0)),\Cal S(\kappa))_-
 \longrightarrow
 H_{\bullet }(\Cal U _{k,n} (\gamma (1)),\Cal S(\kappa))_-
 $$
 be the isomorphism of the Gauss-Manin connection along
 $\gamma$.  Let
 $$
 R_{j,j+1} :
 H^{\bullet} (C^{\bullet}_\lambda
 (V^{\sigma}(q=q(\kappa)),d))
 \longrightarrow
 H^{\bullet}
 (C^{\bullet}_\lambda
 (V^{\sigma\tau }(q=q(\kappa)),d))
 $$
 be the $R$-matrix isomorphism.  Then $ P_{\gamma
 }\cdot\nu_{\*}(\gamma (0),\kappa) = \nu_{\*}(\gamma
 (1),\kappa)\cdot R_{j,j+1}$. \vskip1ex

 \flushpar {\bf {\rm 4.}}  Consider the KZ equation with
 parameter $\kappa$ and with values in {\rm
 Sing}$\,L_\lambda$, see (3.2).  Let {\rm
 Sol}$\,(\kappa)_{\lambda,\sigma}$ be the space of
 solutions to the KZ equation over $D_{\sigma}$.  By
 (4.2.1.1) and (4.2.1.2), for any  $v\in \text{{\rm
 Sing}}\ V^{\sigma} (q=q(\kappa))_\lambda$, the function
 $$
 \psi_{v} : D_{\sigma} \rightarrow \text{{\rm Sing}}\
 L_\lambda,\ z\mapsto \int_{\nu_{\*}(z,\kappa)v}\,N \tag4.2.2
 $$
 is a solution to the KZ equation.  Here $N$ is
 defined by (3.2.3).  Hence, by (4.2.1.1) and (4.2.1.2),
 we have a homomorphism
 $$
 \Cal V _\sigma (\kappa) :
 \text{{\rm Sing}}\  V^{\sigma} (q=q(\kappa))_\lambda
 \to  \text{{\rm Sol}}\
 (\kappa)_{\lambda,\sigma}. \tag4.2.3
 $$
 Let $pr : \text{{\rm Sing}}\ V^{\sigma}
 (q=q(\kappa))_\lambda \to
 \text{{\rm Sing}}\  L^{\sigma} (q=q(\kappa))_\lambda$ be
 the canonical  epimorphism.  We claim that $\ker pr
 \subset \ker \Cal V_{\sigma}(\kappa) $ .  Hence we have a homomorphism
 $$
 \pi_\sigma(\kappa) :
 \text{{\rm Sing}}\  L^\sigma (q=q(\kappa))_\lambda
 \rightarrow \text{{\rm Sol}}\
 (\kappa)_{\lambda,\sigma}.\tag4.2.4$$

 \flushpar {\bf {\rm 5.}}  For an $n$-tree $T$, let $\Cal
 B_{T,\sigma} (\kappa) = \{ v(c,0,T,\sigma,\kappa)\}$
 be the basis in {\rm
 Sing}$\,L^\sigma(q=q(\kappa))_\lambda$ corresponding to
 the tree $T$ with tops marked by
 $m_\sigma(1),\dots,m_\sigma(n)$, see (1.2).  Then for
 any vector $v\in \Cal B_{T,\sigma}(\kappa)$ we have
 $$
 \pi_\sigma(\kappa) (v(c,0,T,\sigma,\kappa))
  = \psi_{c,T,\sigma} .
 $$
 Here $\psi_{c,T,\sigma}$ is the asymptotic solution
 defined by (2.3.1).

 \vskip2ex
 \par {\bf Corollary.} $\pi_\sigma(\kappa)$ {\it is an
 isomorphism.} \vskip2ex

 \flushpar {\bf {\rm 6.}}  Let $u_{T,\sigma}$ be the
 asymptotic zone constructed in (2.2).  If $z\in
 D_\sigma$ and $\{ u_{w,T,\sigma}(z) \}_{w\in {\text{In}}_T}$ are
 small enough, then for any $v(c,0,T,\sigma,\kappa)\in
 \Cal B_{T,\sigma}$ the class
 $\nu_{\*}(z,\kappa)v(c,0,T,\sigma,\kappa)$ sits on the
 $S_k$-orbit of the $t$-critical point $(t^c(z),z)$
 corresponding to the coloring $c$, see (3.6.2).
 Moreover, the asymptotic expansion (3.4.10), (3.4.7.1),
 and (3.3.7.1) of
 $\pi_\sigma(\kappa)v(c,0,T,\sigma,\kappa)$ has the form
 (2.5.8) and properties described in (2.5.7) and (2.5.9).
 \endproclaim

 Theorem (4.2.1) implies (2.4.1), (2.5.7), and (2.5.9).
 Theorem (4.2.1) is proved in (4.4)-(4.6).

\bigskip

 \flushpar {\bf (4.3) Another Form of $U_qs\ell_2$ .}
\bigskip

 Define the $\Bbb Q (q^{\frac{1}{4}})$-algebra
 $\widetilde{U}_q$ as the algebra generated by the
 symbols $E,F,H$ with the relations
 $$\align
 [H,E] &= 2E,\tag4.3.1 \\
 [H,F] &= -2F,\\
 [E,F] &= q^{H/2}-q^{-H/2} .
 \endalign
 $$
 $\widetilde{ U}_q$ has a Hopf algebra structure with the
 comultiplication $\Delta $ defined by
 $$\align
 \Delta (H)  &= H\otimes 1 + 1 \otimes H ,
 \tag4.3.2\\
 \Delta (E)  &= E\otimes q^{H/4} + q^{-H/4} \otimes E , \\
 \Delta (F)  &= F \otimes q^{H/4} + q^{-H/4} \otimes F.
 \endalign
 $$
 Let $\widetilde{ U}_{q=q(\kappa)}$ be the $\Bbb
 C$-algebra obtained from $\widetilde{ U}_q$ by
 specializing the indeterminate $q^{1/4}$ to
 $q(\kappa/4)$.

 For $k>0$, set $(\widetilde{ U}^-)_k = \Bbb CF^k$ and
 $\widetilde{ U}^- = \underset k>0\to{\oplus} \Bbb CF^k$.
 The map $\tilde{\mu} : \widetilde{ U}^- \rightarrow
 \widetilde{ U}^-\otimes \widetilde{ U}^-$ given by (4.1.1),
 in which $f$ is replaced by $F$, defines a coalgebra
 structure.  For a $\widetilde{ U}_{q=q(\kappa)}$-module
 $\tilde{M}$ with highest weights, the map
 $$
 \tilde{\nu} : u\mapsto  \sum_{\ell >0} F^\ell \otimes
 E^{(\ell)} u
 \tag4.3.3
 $$
 defines a $\widetilde{ U}^-$-comodule structure on
 $\tilde{M}$.  For any $\lambda$ define a complex
 $(\widetilde{C}^\bullet_\lambda (\tilde{M}),\tilde{d})$ as
 in (4.1).

 The Hopf algebras $ U _q$ and $\tilde{ U _q}$
 are isomorphic, an isomorphism $\pi : U_q \to \tilde{
 U}_q$ is given by $\pi (h)=H$, $\pi(e)=
 q^{H/4}E$, $\pi (f)= Fq^{-H/4}$.  In particular, these
 formulae give an isomorphism of $ U _{q=q(\kappa)}$
 and $\widetilde{ U}_{q=q(\kappa)}$.

 Assume that $\kappa$ is not a rational number.

 For an integral $m$, let $V(m,q=q(\kappa))$ be the $
 U _{q=q(\kappa)}$ Verma module  with the
highest weight $m$, the generating
 vector $v_m$, and the basis $f^{(\ell )}v_m,\ \ell \geq
 0$.  Let $\widetilde{V}(m,q=q(\kappa)),\ \tilde{v}_m,\
 F^{(\ell )}\tilde{v}_m$ be the corresponding objects
 over $\widetilde{ U}_{q=q(\kappa)}$.  Let
 $$
 \pi_m :
 V(m,q=q(\kappa))\rightarrow\widetilde{V}(m,q=q(\kappa)),\
 f^{(\ell)} v_m \mapsto
 \frac{(Fq^{-H/4})^\ell}{[\ell]_{q(\kappa)}!}
 \tilde{v}_m.
 $$
 Then for any $a\in  U _{q=q(\kappa)}$ and $u\in
 V(m,q=q(\kappa))$ we have $\pi_m(au)= \pi(a) \pi_m(u)$.

 For natural $m_1,\dots,m_n $ and $\sigma\in S_n$ , let
 $$\gather
 V^{\sigma} (q=q(m)) =
 V(m_{\sigma(1)},q=q(\kappa))
 \otimes \dots \otimes
 V(m_{\sigma(n)},q=q(\kappa))\\ \intertext{and}
 \widetilde{V}^{\sigma} (q=q(m)) =
 \widetilde{V}(m_{\sigma(1)},q=q(\kappa))
 \otimes \dots \otimes
 \widetilde{V}(m_{\sigma(n)},q=q(\kappa)) .
 \endgather$$
 For any $k$ and $\lambda$, define a map
 $$\gather
 \Psi :
 C^k_\lambda (V^{\sigma}(q=q(m))) \rightarrow
 \widetilde{C}^k_\lambda (\widetilde{V}^{\sigma} (q=q(m))
 ),\tag4.3.4 \\
 f^{\ell _1} \otimes \dots \otimes f^{\ell _k}
 \otimes u_1\otimes \dots \otimes u_n
 \mapsto
 F^{\ell _1} \otimes \dots \otimes F^{\ell _k}
 \otimes \pi _{m_{\sigma(1)}}
 (u_1) \otimes \dots \otimes \pi _{m_{\sigma(n)}}
 (u_n).\endgather
 $$
 Then
 \vskip2ex \flushpar {\bf (4.3.5)} $\Psi$ {\it defines
 an isomorphism of $C_\lambda^\bullet (V^{\sigma}(q=q(m)))$
 and} $\widetilde{C}_\lambda^\bullet
 (\widetilde{V}^{\sigma}(q=q(m)))$. \vskip2ex

 For a nonnegative integer $m$ denote by
 $\widetilde{L}(m,q=q(\kappa))$ the irreducible
\newline $ \tilde{
 U}_{q=q(\kappa)}$-module with highest weight $m$.  Set
$$\align
 &\text{Sing}\, \widetilde{V}(q=q(\kappa))_\lambda =
 \{ v\in\widetilde{V}^{\sigma} (q=q(\kappa))\,\vert\,
 q^H v=q^\lambda(\kappa) v,\ Ev=0\}, \\ \vspace{2\jot}
 &\text{Sing}\, \widetilde{L}^{\sigma} (q=q(\kappa))_\lambda
 =
 \{ v\in\widetilde{L}^{\sigma} (q=q(\kappa)) \,\vert\, q^H
 v=q^\lambda(\kappa)v,\ Ev=0\}.
 \endalign$$

 \vskip4ex

 \flushpar {\bf (4.4) Construction of the Monomorphism
 $\pmb{\nu}$.}
\bigskip

 Assume that $\kappa,m_1,\dots,m_n,\lambda,k$ are the
 same as in (4.2).

 In \cite{V1} a monomorphism
 $$
 \tilde{\nu} (z,\kappa) : \widetilde{C}^\bullet_\lambda
 (\widetilde{V}^{\sigma}(q=q(\kappa))) \rightarrow \Cal C
 _{k-\bullet}
 (\Cal U _{k,n}(z),\Cal S(\kappa))_-
 \tag4.4.1
 $$
is constructed
 for any $\sigma\in S_n$ and any $z\in D_\sigma$.  Denote
 the image of the monomorphism by $\widetilde{C}_\bullet
 (\Cal U _{k,n}(z),\Cal S(\kappa))_- \subset
\Cal C
 _{\bullet}
 (\Cal U _{k,n}(z),\Cal S(\kappa))_-
$.  It is proved in
 \cite{V1} that the monomorphism has the following four
 properties. \vskip1ex

 \roster
 \item "(4.4.2)"  $\tilde{\nu}(z,\kappa)$ is a
 quasiisomorphism, cf.(4.2.1.1) and
 \cite{V1, \S8.2}. Denote by the same symbol
 $\tilde{\nu}_{\*}(z,\kappa)$ the induced isomorphism in
homology groups.
 \endroster
 \par \vskip1ex \roster\item "(4.4.3)"
 $\tilde{\nu}(z,\kappa)$ is flat with respect to the
 Gauss-Manin connection, cf. (4.2.1.2) and \cite{V1,
 \S8.2}.
 \endroster
 \par \vskip1ex \roster\item "(4.4.4)"
 $\tilde{\nu}_{\*}(z,\kappa)$ transforms the $R$-matrix action
 on $ \widetilde{C}^\bullet_\lambda
 (\widetilde{V}^{\sigma}(q=q(\kappa)))$ into the monodromy action on
$\widetilde{C}_{k-\bullet} (\Cal U _{k,n}(z),\Cal S
 (\kappa))_-$, cf. the precise statement in (4.2.1.3) and
 \cite{V1, \S8.14}.
 \endroster
 \par \vskip1ex \roster\item "(4.4.5)"  For any $v\in
 \text{Sing}\, \widetilde{V}^{\sigma}(q=q(\kappa))_\lambda$,
 the function\endroster

 $$
 \tilde{\psi}_{v} : D_\sigma \to \text{Sing}\,
 L_\lambda,\
 z\mapsto\int_{\tilde{\nu}_{\*}(z,\kappa)v}\,N,
 $$
 \roster \item"${}$" is a solution to the KZ equation.
 By (4.4.3) we have a homomorphism

 $$
 \tilde{\Cal V} _\sigma(\kappa) :
 \text{Sing} \, \widetilde{V}(q=q(\kappa))_\lambda
 \to \text{Sol}\, (\kappa)_{\lambda,\sigma}.
 $$ \vskip1ex

 Let $pr: \text{Sing} \, \widetilde{V}(q=q(\kappa))_\lambda
 \to  \text{Sing} \, \widetilde{L}(q=q(\kappa))_\lambda$ be
 the canonical projection.  It is proved in \cite{V1,
 \S12.2} that $\ker pr\subset\ker
 \tilde{\Cal V}$.  Hence, we have a
 homomorphism \endroster
 $$
 \tilde{\pi}_\sigma :\text{Sing}\,
 \widetilde{L}^{\sigma}(q=q(\kappa))_\lambda
 \to \text{Sol}\, (\kappa)_{\lambda,\sigma}.
 $$

 Let $\Psi : C^k_\lambda
 (V^{\sigma}(q=q(\kappa)))\to \widetilde{C}^k_\lambda
 (\widetilde{V}^{\sigma}(q=q(\kappa)))$ be the isomorphism
 constructed in (4.3.4).  We set
 $$
 \nu(z,\kappa) = (2\pi\kappa)^{^{-\frac{k}{2}}}
 \tilde{\nu}(z,\kappa) \Psi. \tag4.4.6
 $$
 Obviously, this monomorphism has properties
 (4.2.1.1)--(4.2.1.4).  We will prove that the
 monomorphism has properties (4.2.1.5) and (4.2.1.6).

 \vskip4ex
 \subhead (4.5) Proof of Theorem (4.2.1) for
 $\bold{n}\pmb{=}\pmb{2}$ \endsubhead
\bigskip

 For $n=2$ we have
 $$
 \Phi(t,z) = (z_2-z_1)^{^{m_1m_2/\kappa}}
 \prod_{1\leq \ell <j\leq k} (t_\ell -t_j)^{2/\kappa}
 \prod^k_{\ell =1} \ \prod^2_{j=1} (z_j - t_\ell
)^{^{-m_j/\kappa}}. \tag4.5.1
 $$

 There are two permutations in $S_2$:
the identity permutation $id$ and permutation (2,1).  We
 will prove properties (4.2.1.5) and (4.2.1.6) for
 $D_{id}$.  There is only one 2-tree $T$, and therefore,
 there is only one asymptotic zone in $D_{id}$ :
 \newline $u_{T,id} = \{ z_1+z_2,\ u=z_2-z_1\}$.

 Introduce the following new coordinates $x_1,...,x_\ell$ :
$$
t_\ell -z_1= u \cdot x_\ell , \qquad \ell = 1,...,k,
$$
 cf. (3.6.3).  Then
 $$
 \Phi(t(x,u),z(u)) = (-1)^{^{m_1m_2/2\kappa}}
 u^{^{(\frac{m_1m_2}{2}-k(m_1+m_2)+k(k-1))/\kappa}}
 \Phi_{k,m_1,m_2,\kappa} (x).\tag4.5.2
 $$
 where the function $\Phi_{k_1,m_1,m_2,\kappa}$ is given
 by (3.5.1).

 Let $\Delta=\{x\in\Bbb R^k\,\vert\, 0<x_1<\dots
 <x_k<1\}$.

 Let $v_{m_j}\in L(m_j)$ be the generating vector of the
 $s\ell_2$ irreducible module.  Let $\tilde{v}_{m_j}\in
 \widetilde{L}(m_j,\ q=q(\kappa))$ be the generating vector
 of the $\widetilde{ U}_{q=q(\kappa)}$ irreducible
 module.  Let
 $$\gather
  v = \sum^k_{p=0} (-1)^p
 \frac{(m_2-k+1)_{q(\kappa)}\dots
 (m_2-k+p)_{q(\kappa)}}{(m_1)_{q(\kappa)}\dots
 (m_1-p+1)_{q(\kappa)}} \times \tag4.5.3 \\
 \vspace{3\jot}
 \, \qquad q(\kappa)^{^{-p(m_2-2k+p+1)/2}}
 F^{(p)} \tilde{v}_{m_1} \otimes F^{(k-p)} \tilde{v}_{m_2}
 \endgather$$
 be the generating vector of
 Sing$\,\widetilde{L}^{id}_\lambda(q=q(\kappa))$.  Set
 $\gamma (z)= \tilde{\nu}_{\*}(z,\kappa) v\in H_k(\Cal U
 _{2,k}(z),\Cal S(\kappa))_-$ for $z\in D_{id}$.

 According to the explicit construction of
 $\nu(z,\kappa)$ in \cite{V1}, we have
 $$\multline
\int_{\gamma (z)} N(t,z) = \\ \vspace{3\jot}
 q(\kappa)^{^{-\frac{m_2}{4}+\frac{k(k-1)}{4}}}
 (m_2)_{q(\kappa)} \dots
 (m_2-k+1)_{q(\kappa)} \cdot
 u^{^{(\frac{m_1m_2}{2}-k(m_1+m_2)+k(k-1))/\kappa}}\times \\
\vspace{3\jot} k!
\int_{\Delta}\ \prod^{k}_{j=1} x_j^{-m_1/\kappa}
 (1-x_j)^{-m_2/\kappa} \prod_{1\leq j<\ell \leq k}
 (x_\ell -x_j)^{2/\kappa} A(x)
 dx_1\wedge \dots \wedge dx_k  \cdot f^\ell v_{m_1} \otimes
 f^{k-\ell}v_{m_2} ,
 \endmultline\tag4.5.3$$
and
$$
 A_\ell (x) = \sum_{\sigma\in S(k;\ell ,k-\ell)}
\, \prod^k_{i=1} \frac{1}{x_i-y_{\sigma(i)}},$$
 where $y_1=0,\ y_2=1$, and $S(k;\ell ,k-\ell )$ is
 defined in (3.2.2).

 The value of the integral over $\Delta $ is considered
 in the sense of analytic continuation from the domain of
 parameters where $\kappa$ is positive and $m_1,m_2$ are
 negative.

 This formula is the decisive property of
 $\tilde{\nu}(z,\kappa)$ which allows us to prove
 (4.2.1.5) and (4.2.1.6).

 Consider the single basic vector $v(c,0,T,id,\kappa)\in
 \Cal B_{T,id}(\kappa)$ of the one dimensional space
 Sing$\,L^{id}_\lambda(q=q(\kappa))$.  (4.4.6) implies
 that
 $$\multline
\int_{\nu(z,\kappa)v(c,0,T,id,\kappa)}
 N = (2\pi\kappa)^{-k/2} q(\kappa)^{ -\frac{m_2}{2} +
 \frac{3}{4}k(k-1)} \times \\ \vspace{3\jot}
 u^{^{(\frac{m_1m_2}{2}-k(m_1+m_2)+k(k-1))/\kappa}}
 k! \, I_k(m_1,m_2;\kappa)\times \\ \vspace{2\jot}
\prod^{k-1}_{j=0}
 \biggl( \biggl( \pmb{1} - \frac{\kappa}{m_1+m_2-2k+j+2}
 \biggr)
 (m_2-j)_{q(\kappa)}\biggr)\cdot \{ v_1 \ ,_{k}\ v_2 \}
 \endmultline\tag4.5.4
 $$
 where the vector $\{ v_1\ ,_{k}\ v_2  \}\in $
 Sing$\,(L(m_1)\otimes L(m_2))_\lambda$ is given by
 (1.4.1) and $I_k(m_1,m_2;\kappa)$ is the Selberg
 integral given by (1.5.1).  By (1.5.2) and (1.5.3), we
 have
 $$
\int_{\nu_{\*}(z,\kappa)v(c,0,T,id,\kappa) }\
 N =
 C(c,T,\kappa) \cdot
 u^{^{(\frac{m_1m_2}{2}-k(m_1+m_2)+k(k-1))/\kappa}} \{
 v_1\ ,_k\ v_2 \}.
 \tag4.5.5
 $$

 The right hand side is exactly the asymptotic solution
 defined by (2.3.1).  This proves (4.2.1.5).

 Now assume that $\kappa=is,\ s\in \Bbb R$, and $s\to
 +0$.  We compute the asymptotic expansion of the right
 hand side of (4.5.4) using (1.5.1) and the Stirling
 formula.  This shows that the asymptotic expansion of
 $\pi_{id}(\kappa)v(c,0,T,id,\kappa)$ has the form
 (2.5.8) and the properties described in (2.5.7) and (2.5.9).

 According to the construction in \cite{V1}, the class
 $\nu(z,\kappa)v(c,0,T,id,\kappa)$ is flat with respect
 to $\kappa$, see (3.4).  The function $\Phi(t,z)$ has
 exactly one $S_k$-orbit of $t$-critical points, see
 (3.5).  To show that the class sits on this orbit we
 need the following lemma.

 \proclaim{(4.5.6) Lemma}  Let $[\gamma (\kappa)]\in
 H_k(\Cal U _{k,2}(z),\Cal S(\kappa))$ be a class flat
 with respect to $\kappa$.  Then $[\gamma (\kappa)]$ sits
 on the orbit of $t$-critical points of $\Phi(t,z)$ or
 for any $P>0$ there exist a chain
 $$
 \gamma (\kappa) =  \sum^M_{j=1} (c_j,\alpha_j(\kappa))
 $$
 representing $[\gamma (\kappa)]$, flat with respect to
 $\kappa$, and such that, for every $j=1,\dots,m$ ,  we have
 $$
 \lim \Sb \kappa=is\\s\to +0 \endSb q(\kappa)^{^{P}}
 \sup_{c_j}
 \vert \alpha_j(\kappa) \vert =0. \tag4.5.7
 $$
 \endproclaim

 \proclaim{(4.5.8) Corollary} If $[\gamma (\kappa)]$ does
 not sit on the $\Cal S_k$ orbit then
 $$
 \lim \Sb \kappa=is\\s\to +0 \endSb q(\kappa)^{^{P}}
 \int_{[\gamma (\kappa)]}N =0
 $$
 for all $P$.
 \endproclaim

 \demo{Proof}  First, remark that $\Phi^{\kappa}$ is a
 rational function on $\Cal U _{k,2}(z)$.

 Now consider the compactification of $\Cal U _{k,2}(z)$
 in $(\Bbb CP^1)^k$.  Let $X\subset(\Bbb CP^1)^k$ be the
 divisor of singularities of $\Phi$.  A resolution of
 singularities of $X\subset(\Bbb CP^1)^k$  is a proper
 analytic map $F: (A\subset B)\to (X\subset (\Bbb
 CP^1)^k)$ such that $A$ is a divisor with normal
 crossings
 in a nonsingular$B$ and
$F\big\vert_{B-A} : B-A \to (\Bbb CP^1)^k-X$ is a
 biholomorphism.  The condition $k\leq \min (m_1,m_2)$
 implies that there exists a resolution having the
 following property:

 \roster
\item"{(4.5.9)}" For any point $a\in A$, there exist local
 coordinates $u_1,\dots,u_k$ on $B$ centered at $a$ and
 such that $A$ is defined by equation
$$u_1 \cdot \dots \cdot u_r=0$$  for some $r\in \{
 1,\dots,k\}$ and $$\Phi^{\kappa}\circ F(u) = u^{\ell
 _1}_1 \cdot \dots \cdot u^{\ell _r}_r$$ where $\ell _1\cdot\dots\cdot\ell
 _r\neq 0$.
\endroster
 \vskip1ex

 Let $S=\kappa\ln \Phi$.  Then $S= \sum^r_{j=1} \ell
 _j(\ln\vert u_j \vert +i \arg u_j)$.  Introduce the new
 real coordinates $v_j = \ln\vert u_j \vert$ and $w_j=\arg
 u_j$.  Let $H_1 , \dots , H_r$ be arbitrary real numbers
 such that $H_1\ell _1 + \dots + H_r\ell _r = - 1 .$  Set $Y=
 \sum_j H_j \frac{\partial}{\partial w_j}$.  The vector
 field $Y$ decreases Im $S$ in a neighborhood of
 the point $a$.

  To prove the lemma we start with an
 arbitrary representing chain $\gamma (\kappa)$, flat
 with respect to $\kappa$.  Then we deform $\gamma
 (\kappa)$ into the direction of decrease of Im$\,S$.  To
 deform the chain we use local fields $\{ Y \}$
 constructed above in a neighborhood of $X$ and we use the
 field grad(Im $S$) in the \lq finite\rq part of
 the $(\Bbb CP^1)^k\backslash X$.  This procedure will
 push the chain onto the orbit of $t$-critical points or
 will create a new representing chain with property
 (4.5.7).

 The lemma and Theorem (4.2.1) for $n=2$ are proved.
 \enddemo

 \vskip4ex

 \subhead (4.6) Proof of Theorem (4.2.1) \endsubhead
\bigskip

 To simplify notation we'll prove the theorem for the
 case $n=3,\ \sigma=id\in S_3$, and for the 3-tree shown
 in (4.6.1).  The general case is completely similar.
\newline \ \newline \ \newline \ \newline \ \newline
\newline \ \newline \ \newline \ \newline \ \newline
$$
 \tag4.6.1
 $$

 In this case the asymptotic zone has the form $u=
 u_{T,id} = \{ z_1+z_2+z_3,u_1,u_2\}$ where
 $z_3-z_2=u_2,\ z_2-z_1=u_1u_2$.  If $u_1$ and $u_2$
 are small, then $z_2-z_1\ll z_3-z_2$.

 Let the tops of $T$ be marked by nonnegative integers
 $m_1,m_2,m_3$ and let $c$ be an admissible coloring of
 $T$ having level $\lambda = m_1+m_2+m_3-2k$.  Set
 $c_1=c(w_1),\ c_2= c(w_2)$,  then $c_1+c_2=k$.

 Let $v_m$ be the generating vector of the $s\ell_2$
 module $L(m)$.  Let $\overline{v}_m$ be the generating
 vector of the $ U _{q=q(\kappa)}$ module
 $L(m,q=q(\kappa))$.

 Set $v=\{ \{ v_{m_1} \ ,_{c_1}\ v_{m_2}\},_{c_2}\ v_{m_3}\}$,
$\overline{v}=(
 ( \overline{v}_{m_1} \ ,_{c_1}\
 \overline{v}_{m_2})\ ,_{c_2}\ \overline{v}_{m_3})$, see
 notations in (1.2.1) and (1.4.1).

 Consider the solution to the KZ equation given by
 $$
 \psi :z\mapsto \int_{\nu(z,\kappa)\overline{v}} N,
 \tag4.6.2
 $$
 where $z=(z_1,z_2,z_3),\ z_1<z_2<z_3$.  Our problem is
 to show the following statements (4.6.3) and (4.6.4).
 $$\gather
 \psi(z(u)) = C(c,T,\kappa)\cdot  u_1^{\mu_1/\kappa}
 \,u_1^{\mu_2/\kappa}(v+\Cal O(u)),\tag4.6.3\\ \vspace{2\jot}
 C(c,T,\kappa) = \frac{k!}{c_1!c_2!} \cdot
 J_{c_1}(m_1,m_2;\kappa) \cdot
 J_{c_2}(m^1,m_3;\kappa)
 \endgather$$
 where $m^1= m_1+m_2-2c_1$, see (1.5.3), and the real
 numbers $\mu _1,\mu _2$ are defined in (2.3.1).

 \proclaim{(4.6.4)} Assume that $\kappa=is, s\to +0$.
 Then for small $u_1,u_2$ the class $\nu
 (z,\kappa)\overline{v}$ sits on the orbit of the
 $t$-critical point $(t^c(z),z)$ such that
 $$\gather
 t^c_p(z) -z_1 = (a_p +\Cal O(u)) u_1u_2,\qquad p=1,\dots
 ,c_1 , \\ \vspace{2\jot}
 t^c_{p+c_1} (z) -z_2 = (b_1 +\Cal O(u)) u_2,\qquad
 p=1,\dots,c_2 ,
 \endgather$$
 where $( a_1,\dots,a_{c_1})$ (resp. $(b_1,\dots
 ,b_{c_2}))$ is a critical point of
 $\Phi_{c_1,m_1,m_2,\kappa}$ (resp.
 $\Phi_{c_2,m^1,m_3,\kappa}$), see {\rm (3.6.2)}.  Moreover,
 the asymptotic expansion of $\psi(z(u))$ as $s\to +0$
 has the form described in {\rm (2.5.6)} and the properties
 described in {\rm (2.5.7)} and {\rm (2.5.9)}.
 \endproclaim

 \demo{Proof}  To prove these statements we will choose a
 special chain representation for the class
 $\nu(z,\kappa)\overline{v}$.  We will use the
 construction of iterated cycles described in \cite{V1,
 \S14}.

 Consider the space $\Bbb C^{2+c_1}$ with coordinates
 $z_1,z_2,t_1,\dots,t_{c_1}$.  Let
 $$\gather
 pr_I : \Bbb C^{2+c_1} \rightarrow \Bbb C^2, \
 (z_1,z_2,t) \mapsto (z_1,z_2),\\
  \Phi_I = (z_2 - z_1)^{\frac{m_1m_2}{2\kappa}}
 \prod_{1\leq \ell <j\leq c_1} (t_\ell -t_j)^{2/\kappa}
 \cdot
 \prod^k_{\ell =1}\ \prod^2_{j=1} (z_j - t_\ell
 )^{-m_j/\kappa}.
 \endgather$$
 By (4.4.6) we have a map
 $$\multline
 \nu_I (z_1,z_2,\kappa) : \text{Sing}\,
 (L(m_1,q=q(\kappa))
 \otimes L(m_2,q=q(\kappa)) )_{m^1=m_1+m_2-2c_1}
 \longrightarrow \\\vspace{3\jot} H_{c_1}(pr_I (z_1,z_2)^{-1},\Cal
 S_I(\kappa))_-,\endmultline
  $$
 where $z_1,z_2\in \Bbb R$, $z_1<z_2$, and $\Cal
 S_I(\kappa)$ is the local system defined by $\Phi_I$.

For $z_1=0,\ z_2=1$, fix a chain representative $\gamma
 _I$ for the class $[ \gamma _I(0,1)] =
 \nu_I (0,1,\kappa)(\overline{v}_{m_1}\ ,_{c_1}
\  \overline{v}_{m_2})$ , flat with respect to $\kappa$, cf.
 (3.4.1).
 \enddemo

 For fixed numbers $z^0_1, z^0_2$ the map
 $$
 T:(z_1,z_2,t_1,\dots,t_{c_1}) \mapsto
 ( (z^0_2 - z^0_1) z_1+z^0_1,
 (z^0_2 - z^0_1) z_2+z^0_1,
 (z^0_2 - z^0_1) t_1+z^0_1, \dots )
 $$
 sends $pr^{-1}_I(0,1)$ to $p^{-1}_I (z^0_1,z^0_2)$.
 Hence, the image of $\gamma _I$ under $T$ gives a
 representative
 $$
 \gamma _I (z^0_1,z^0_2)
 = \sum^{M_I}_{j=1}
 (C^I_j (z^0_1,z^0_2),\alpha^I_j(\kappa))
 \tag4.6.6
 $$
 of the class $\nu_I (z^0_1,z^0_2,\kappa)
 (\overline{v}_{m_1}\ ,_{c_1}\ \overline{v}_{m_2})$ for
 arbitrary $z^0_1<z^0_2$.  This representation is flat
 with respect to $\kappa$.

 Similarly, consider the space $\Bbb C^{2+c_2}$ with
 coordinates $z_2,z_3,t_{c_1+1},\dots,t_k $.  Let
 $$
 pr_{II}: {\Bbb C}^{2+c_2} \to \Bbb C^2, \qquad (z_2,z_3,t)\mapsto
(z_2,z_3),
$$
$$
\Phi_{II} =  (z_3 - z_2)^{^{m^1m_3/2\kappa}}
 \prod_{c_1 < \ell <j\leq k}
 (t_\ell -t_j)^{2/\kappa}
 \prod^k_{\ell =c_1+1}
 (z_2 - t_\ell)^{^{-m^1/\kappa}}
 (z_3 - t_\ell)^{-m_3/\kappa} .
 $$
 By (4.4.6) we have a map
 $$\multline
 \nu_{II} (z_2,z_3,\kappa) : \text{Sing}(L(m^1,q=q(\kappa)) \otimes
 L(m_3,q=q(\kappa))_{_{m^1+m_3-2c_2}}
 \rightarrow\\ \vspace{2\jot}
 H_{c_2} (pr_{II}^{-1}(z_2,z_3) , \Cal S_{II}(\kappa))_-
 \endmultline$$
 where $z_2,z_3 \in \Bbb R,\ z_2<z_3$, and $\Cal
 S_{II}(\kappa)$ is defined by $\Phi_{II}$.  As above,
 construct a chain representation
 $$
 \gamma _{II} ( z_2,z_3 ) =
 \sum^{M_{II}}_{\ell =1}
 (C^{II}_\ell ( z_2,z_3 ),\alpha^{II}_{\ell }(\kappa) )
 \tag4.6.7
 $$
 of the class $\nu_{II}( z_2,z_3 )
 (\overline{v}_{m^1},_{c_2}\ \overline{v}_{m_3})$ which is
 flat with respect to $\kappa$.

 Using the
chains $\gamma _I,\gamma _{II}$, we will construct
 a chain representing \newline $\nu(z,\kappa)
 ((\overline{v}_{m_1} ,_{c_1}\ \overline{v}_{m_2}),_{c_2}
\ \overline{v}_{m_3} )$.

 Let $u_1,u_2$ be small, then
 $z_2-z_1\ll z_3-z_2$.

 Identify $pr{^{-1}_I}
 (z_1,z_2)$ with $\Bbb C^{c_1}$ using the coordinates
 $t_1,\dots,t_{c_1}$ ,
identify $pr^{-1}_{II} (z_2,z_3)$ with $\Bbb
 C^{c_2}$ using the coordinates $t_{c_1+1},\dots,t_k$, and
 identify $pr^{-1}_{3,k}(z_1,z_2,z_3)$ with $\Bbb C^k= \Bbb
 C^{c_1} \times \Bbb C^{c_2}$.

Using these identifications,
 for any $j=1,\dots ,M_I,\ \ell =1, \dots , M_{II}$,
 define a $k$-cell $C_{j\ell }(z_1,z_2,z_3)$ in
 $pr^{-1}_{3,k} (z_1,z_2,z_3):$
 $$
 C_{j\ell} (z_1,z_2,z_3) =
 C_j (z_1,z_2) \times C_\ell (z_2,z_3).
 $$

 Using $\alpha^{II}_j$ and $\alpha^{II}_\ell $ we will
 define a coefficient of $C_{j\ell} $ in the local system
 $\Cal S(\kappa)$ defined by the function $\Phi$ described
 in (3.1.1) for $n=3$.

 $\alpha^I_j$ has the form $\epsilon^I_j\cdot\beta ^I_j$
 where $\epsilon^I_j$ is a rational function of
 $q(\kappa)^{\frac14}$ and $\beta ^I_j$ is a branch of
 $\Phi_I$ over $C^I_j$.  $\alpha^{II}_\ell $ has the form
 $\epsilon^{II}_{\ell}\cdot\beta ^{II}_\ell$ where
 $\epsilon^{II}_\ell$ is a rational function of
 $q(\kappa)^{\frac14}$ and $\beta ^{II}_\ell$ is a branch of
 $\Phi_{II}$ over $C^{II}_\ell$.  We define the coefficient of
 $C_{j\ell }$ in ${\Cal S}(\kappa)$ by
 $$
 \alpha_{j\ell } = \epsilon^I_j\cdot\epsilon^{II}_\ell
\cdot \beta _{ij}
 $$
 where $\beta _{ij}$ is the branch of $\Phi$ over $C_{j\ell
 }$ defined below.

 Any branch of $\Phi_I,\Phi_{II}$, or $\Phi$
 is defined by determining arguments of all differences
 $t_\ell -t_j$, $z_j - t_\ell$, $z_\ell -z_j$ in the
 formulae for $\Phi_I, \Phi_{II}$, and $\Phi$.
 Therefore, we assume that over $C^I_j$ (resp.
 $C^{II}_\ell$) argument of every difference in
 $\Phi_I$ (resp. $\Phi_{II}$) is fixed.  Let us determine
 argument of every difference in $\Phi$ by the
 following rule.

 Choose an argument of $z_3-z_2,\ z_2-z_1,\ z_b - t_a$
(for  $a>c_1,\ b=2,3$), $z_b - t_a$ (for $a<c_1$ and $ b=1,2$),
 $t_a - t_b$
 (for $1\leq a<b\leq c_1$ and for $c_1<a<b\leq k$) the same
 as those in $\beta ^I_j$ and $\beta ^{II}_\ell$.  The
 function $z_1 - t_a$ for $a>c_1$ is approximately equal to
 $z_2 - t_a$ on $C_{j\ell }$.  Choose the argument of
 $z_1 - t_a$ which is close to the argument of $z_2 - t_a$ in
 $\beta ^{II}_\ell $.

 The function $z_3 - t_a$ for $a\leq c_1$ is approximately
 equal to $z_3-z_2$ on $C_{j\ell }$.  Choose the argument
 of $z_3 - t_a$ which is close to the argument of $z_3-z_2$ chosen
 in $\beta ^{II}_\ell $.

 The function $t_a-t_b$ for $a\leq c_1$ and $b>c_1$ is
 approximately equal to $z_2-t_b$ on $C_{j\ell }$.
 Choose the argument of $t_a-t_b$ which is close to the
 argument of $z_2-t_b$ on $C_{j\ell}$.

 It is easy to see that
 $$
 \tau (z_1z_2z_3)=
 \sum^{M_I}_j\ \sum^{M_{II}}_\ell
 (C_{j\ell },\alpha_{j\ell })
 $$
 is a cycle.  The group $S_k$ acts on the space of chains
 with coefficients in $\Cal S(\kappa)$.  Set
 $$
 \gamma (z_1,z_2,z_3) = \frac{1}{c_1!c_2!}
 \sum_{\chi \in S_k} (-1)^{^{\vert \chi \vert}}
 \chi \cdot \tau (z_1,z_2,z_3).
 $$

 \vskip1ex
 According to \cite{V1, \S14} we have
\roster
\item "{ (4.6.8)}" {\it The chain $\gamma
 (z_1,z_2,z_3)$ represents the class
 $\nu(z_1,z_2,z_3)\overline{v}$.}
\endroster
 \vskip1ex

 Now introduce the new coordinates $x_1,\dots ,x_k$ in
 $pr^{-1}_{3,k} (z_1,z_2,z_3)$:
 $$
 \aligned
 &t_p-z_1 =x_pu_1u_2 , \\
 &t_p-z_2 =x_pu_2 ,\endaligned \quad \aligned &p=1,\dots,c_1\\
 &p=c_1+1,\dots ,k.
 \endaligned
 $$

 The cells $C_{j\ell }$ of the chain $\tau$ written in
 coordinates $x_1,\dots ,x_p$ do not depend on $u_1,u_2$.
 The dependence of $\Phi$ in those coordinates has the
 form
 $$
 \Phi = u_1^{^{\mu_1 /\kappa}} u_2^{^{\mu_2 /\kappa}}
 (\Phi_{_{c_1,m_1,m_2,\kappa}}
 (x_1,\dots,x_{c_1})\cdot
 \Phi_{_{c_2,m^1,m_3,\kappa}}
 (x_{c_1+1},\dots,x_k)
 +\Cal O(u)).\tag4.6.9
 $$
 Knowing that $\gamma _I(0,1)$ sits on $(a_1,\dots
 ,a_{c_1})$ and $\gamma _{II}(0,1)$ sits on $(
 b_1,\dots ,b_{c_2} )$ we may conclude that $\gamma $
 sits on the $t$-critical point described in (4.6.4).
 Knowing that integrals over $\gamma _{I}$ and $\gamma
 _{II}$ have the  desired asymptotic expansion, see
 (4.5), we may conclude that the asymptotic expansion of
 the function $\psi$ has the form described in (2.5.6)
 and the properties described in (2.5.7).
 Formula (4.6.9) also shows that the integral
$$\int_{\gamma
 (z_1,z_2,z_3)} N$$
 is the asymptotic
 solution $\psi_{c,T,id}$.

 Theorem (4.2.1) is proved.

\newpage

\centerline{\bf References}
\bigskip

[AGV] V.I.Arnold, S.M.Gusein-Zade, A.N.Varchenko, Singularities
of Differentiable Maps, {\bf 2}, {\it Birkhauser}, 1988.

[B] H. Bethe, {\it Z. Phys.} {\bf 71} (1931), 205.

[Ba] H.M.Babujian, Off-shell Bathe ansatz equation and N-point
correlators in the SU(2) WZNW theory, preprint Bonn-HE-93-22.

[BF] H.M.Babujian, R.Flume, Off shell Bethe ansatz equation
for Gaudin magnets and solutions of Knizhnik-Zamolodchikov
equations, preprint Bonn-HE-93-30.

[D] V.Drinfeld, Quasi-Hopf algebras, {\it Leningrad Math. J.} {\bf
1} (1989), no. 6.

[FFR] B.Feigin, E.Frenkel, N.Reshetikhin,
Gaudin model, Bethe ansatz and correlation functions at the
critical level, preprint (1994), 39 pages.

[FT] L.Faddeev, L.Takhtajan, {\it Uspehi Mat. Nauk} {\bf 34},
no. 5 (1979), 13-63.

[FW] G.Felder, C.Wieczerkowski, Topological representations of $U_q
(s\ell_2)$, {\it Comm. Math. Phys.} {\bf 138} (1991) 583-605.

[G] M. Gaudin, Diagonalizations d'une classe
d'hamiltoniens de spin, {\it Jour. de Physique} {\bf 37}, no. 10
(1976), 1087--1098.

[K] M.Kashiwara, Crystalizing the $q$-Analogue of Universal
Enveloping Algebras, {\it Comm. Math. Phys.}{\bf 133} (1990) 249-260.

[Ko] T.Kohno, Monodromy representations of braid groups and Yang-Baxter
equations, {\it Ann. Inst. Fourier} {\bf 37} (1987), 139-160.

[KR] A.Kirillov, N.Reshetikhin, Representations of the algebra $U_q
(s\ell(2)) , q-$orthogonal polynomials and invariants of links,
preprint, Leningrad (1988), 55 pages.

[KZ] V. Knizhnik and A. Zamolodchikov, Current algebra and
Wess-Zumino models in two dimenisons, {\it Nucl. Phys.} {\bf B 247}
(1984), 83--103.

[M] M.Mahta, Random Matrices, {\it Acad. Press}, 1991.

[R] N.Reshetikhin, Jackson type integrals, Bethe vectors, and
 solutions to a difference analog of the Knizhnik-Zamolodchikov system,
{\it Lett. Math. Phys.}{\bf 26} (1992), 153-165.

[Sz] G.Szego, Orthogonal Polynomials, AMS, 1939.

[TV] V. Tarasov and A. Varchenko, Jackson integral
representations for solutions of the KZ equation,  preprint:
RIMS--949 (1993).

[SV] V. Schechtman and A. Varchenko, Arrangements of
hyperplanes and Lie algebra homology, {\it Invent.  Math.} {\bf 106}
(1991), 139--194.

[V1] A. Varchenko, Multidimensional hypergeometric functions
and representation theory of Lie algebras and quantum groups,
preprint (1992), 401 pages.

[V2] A. Varchenko, Critical points of the product of powers
of linear functions and families of bases of singular vectors,
preprint (1993), 15 pages.

\enddocument